\documentclass[10pt,journal,compsoc]{IEEEtran}
\usepackage{xargs}
\usepackage{todonotes}

\usepackage{marginnote}

\usepackage{subfig}
\usepackage[export]{adjustbox}

\usepackage{tikz,pgf, ifthen}
\usetikzlibrary{backgrounds}
\usetikzlibrary{calc, arrows,chains,positioning,scopes}

\usepackage{algorithm}
\usepackage{algpseudocode}

\usepackage{placeins}

\usepackage{csvsimple}
\usepackage[round-mode=places, round-integer-to-decimal, round-precision=2,group-separator={,},group-four-digits = true]{siunitx}
\let\oldnum\num
\renewcommand{\num}[1]{\oldnum[group-separator={,},group-four-digits = true,round-mode=off]{#1}}

\usepackage{booktabs}   %% For formal tables:
\usepackage[utf8]{inputenc}

\newcommand{\code}[1]{\texttt{#1}}

\ifCLASSOPTIONcompsoc
  \usepackage[nocompress]{cite}
\else
  \usepackage{cite}
\fi
\ifCLASSINFOpdf
\else
\fi
\usepackage{amsmath}
\usepackage{amsfonts}
\usepackage{amsthm}

\theoremstyle{definition}
\newtheorem{definition}{Definition}[section]

\usepackage[capitalise]{cleveref}

\begin{document}
\title{Work-stealing prefix scan: Addressing
load imbalance in large-scale image registration}

% FIXME
\author{Marcin~Copik$^1$,
        Tobias~Grosser$^2$,
        Torsten~Hoefler$^1$,~\IEEEmembership{Member,~IEEE,}
        Paolo~Bientinesi$^3$,
        Benjamin~Berkels$^4$\\
$^1$Department of Computer Science, ETH Zurich; $^2$School of Informatics, University of Edinburgh;\\
$^3$Department of Computing Science, Umeå University; $^4$AICES, RWTH Aachen University
}

\IEEEtitleabstractindextext{%
\begin{abstract}
Parallelism patterns (e.g., map or reduce) have proven to be effective tools
for parallelizing high-performance applications.
In this paper, we study the recursive registration of a series of electron
microscopy images -- a time consuming and imbalanced computation necessary for nano-scale
microscopy analysis. We show that by translating the image registration  into a specific instance of the prefix scan, we can convert this seemingly sequential problem into a parallel computation that scales to over thousand of cores.
We analyze a variety of scan algorithms that behave similarly for common
low-compute operators and propose a novel work-stealing procedure for a hierarchical prefix scan. 
Our evaluation shows that by identifying a suitable and well-optimized prefix scan algorithm, we reduce time-to-solution
on a series of 4,096 images spanning ten seconds of microscopy
acquisition from over 10 hours to less than 3 minutes
(using 1024 Intel Haswell cores), enabling derivation of material properties at
nanoscale for long microscopy image series.
\end{abstract}

%\begin{IEEEkeywords}
%Computer Society, IEEE, IEEEtran, journal, \LaTeX, paper, template.
%\end{IEEEkeywords}
}

\maketitle

\IEEEdisplaynontitleabstractindextext

% \fi
\IEEEpeerreviewmaketitle

\IEEEraisesectionheading{\section{Introduction}\label{sec:introduction}}
Many seemingly sequential algorithms in which the computation of element
$x_{i+1}$ depends on element $x_i$ can be parallelized with a \emph{prefix
scan} operation. 
Such an operation takes a binary and associative operator $\odot$ and an input array
$x_0, x_1, \ldots, x_n$ and produces the output array $y_0, y_1, \ldots,
y_n$. Every element $y_i$ in the output array is the result of the binary
and inclusive combination of all ``previous'' elements and the current one in the input array: $y_i =
 x_0 \odot x_1 \odot \cdots \odot x_i$. 
An \emph{exclusive prefix scan} computes an output array where each combination does
not include the input element with the corresponding index $y_i =
x_0 \odot x_1 \odot \cdots \odot x_{i-1}$. The difference between the two variants can be easily compensated since both results are always one shift away. A transformation from an exclusive to an inclusive sum requires a shift of the result by one position to the left and one operator application to produce last element.
%This defines an \textit{inclusive} scan where each new element with index $i$ is a sum of first $i$ elements of input sequence. A less popular variation is known as an \textit{exclusive} prefix sum where the new $i$-th value is a sum of first $i-1$ input values
%\begin{align*}
%y_{n-1} \: &= \: x_{1} \odot x_{2} \odot \dots \odot x_{n-2} = \odot_{j=1}^{n-2} x_{j}
%\end{align*}
%Inclusive prefix scan algorithms are far more common but the exclusive scan is a fundamental basic block used in the distributed prefix sum, as defined in Section~\ref{sec:distributed_prefix_sum}. The difference between two variants can be easily compensated since results are always one shift away. A transformation from an exclusive to an inclusive result requires a shift of results by one position to the left and computation
%
One can show that, if a sequentially-dependent transformation consuming
$x_i$ and producing $x_{i+1}$ can be expressed as such a binary
 operator, then the problem can be parallelized
using a prefix scan.

\begin{figure}[tb]
		\includegraphics[width=\dimexpr0.49\textwidth]{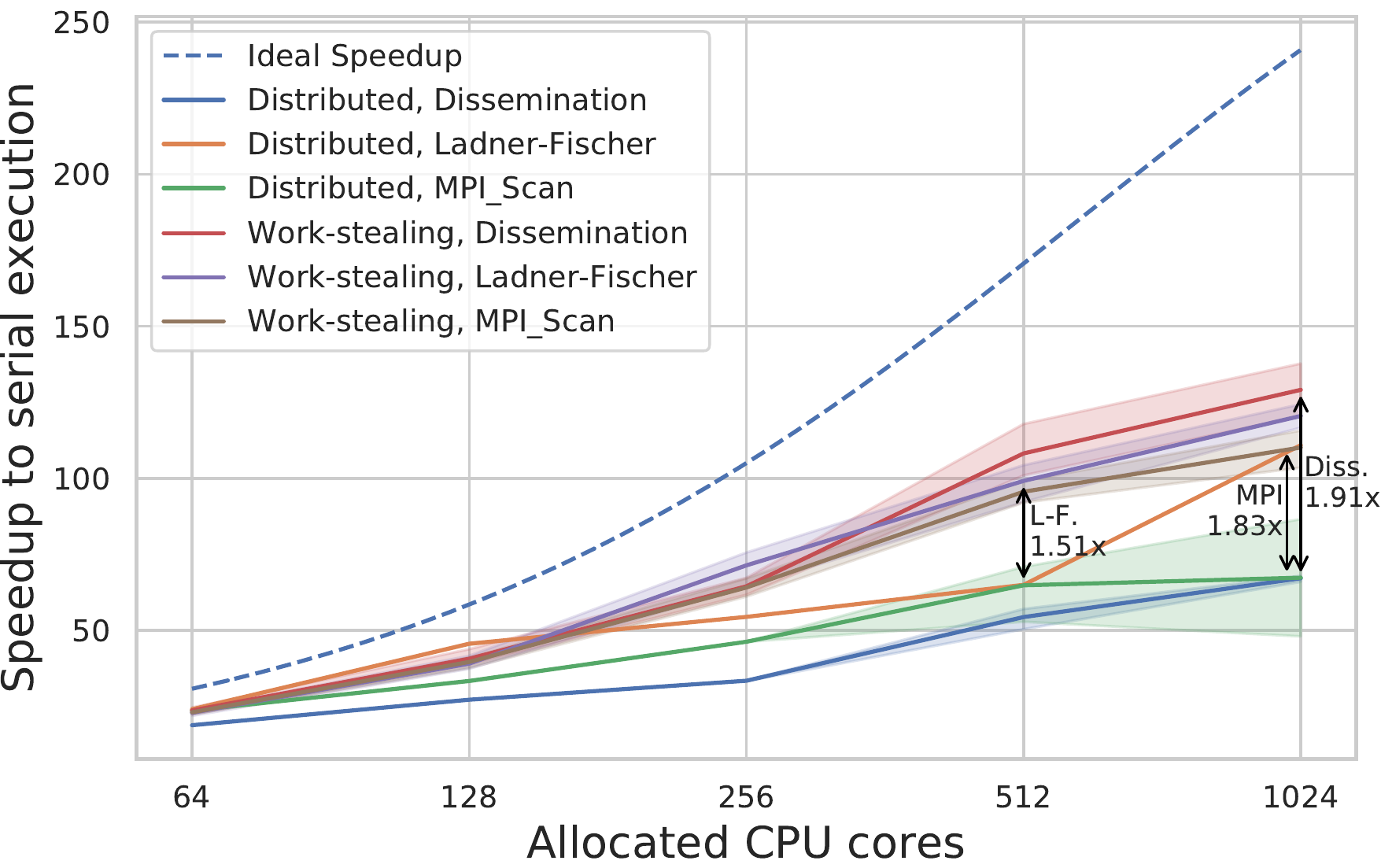}
	\centering
  \caption{The strong scaling of distributed prefix scan (Section~\ref{sec:distributed_prefix_scan}) on the \textit{scan} image registration for three
    variants of prefix scan algorithms (Section~\ref{sec:prefix_scan}). Experimental results (solid lines) were obtained on the Piz Daint~\cite{pizdaint} system for $\num{4096}$ images. Theoretical bound~\eqref{eq:scan_speedup} is discussed in Section~\ref{sec:strong_scaling}.}
	%\subfloat[The strong scaling of prefix scan image registration.]{
	
	%\resizebox*{0.9\width}{0.9\totalheight}{
	%}
	\label{fig:strong_scan2}
	%\vspace{-9mm}
\end{figure}

This powerful construct has numerous uses in parallel computing. 
It enables parallelization of multiple non-trivial problems that might
seem to be inherently sequential, including finite state machines,
solving linear tridiagonal systems, parallelization of many sequential loops
with dependencies, or sequential chains of computations that can be
modeled as a function
composition~\cite{Chatterjee:1990:SPV:110382.110597,BlellochTR90,Cormen:2001:IA:580470,Blelloch:1990:VMD:91254}.
However, the sequentiality in the original problem causes some
overheads---the resulting parallel algorithm is either highly parallel
or work efficient but not both at the same time. The
more processes can be used effectively, the more additional work (i.e.,
applications of $\odot$) has to be performed in parallel.
If the workload is balanced and all processes run at the same speed,
then the additional work does not delay the processing. In this
scenario, a prefix scan is as fast as a simple reduction albeit with a
higher energy consumption due to additional computation. 

In this work, we parallelize an application in the area of large-scale
image registration in electron microscopy. Given the importance of
microscopy data for analysis of material properties and temporal
changes at nanoscale, a critical objective is to enable the processing of very long
sequences of microscopy images by a domain specialist without running
days long computations. We look at the grand scheme of registration and represent
this seemingly sequential procedure as a composition of two steps, a massively
parallel preprocessing phase and a prefix scan.
In contrast to most other applications of prefix scans, our application requires \emph{expensive
	and highly load-imbalanced operators with nearly trivial communication}. 
We are first to show (1) how the communication pattern in near-optimal
circuits by Ladner and Fischer can be tuned for MPI execution on
large-scale compute clusters and (2) we develop a novel node-local
work-stealing algorithm for general prefix scans, found in a large
variety of recursive and seemingly sequential computations.
The load balancing scan enables the parallelization of problems that would otherwise
be considered inefficient given imbalanced computation, sparse iteration space,
and a tightly constrained form of prefix scan.

We apply the load balancing prefix scan to the registration procedure 
and we show that the performance of distributed prefix scans can be significantly improved
even for a highly imbalanced application. In the strong scaling experiment (Section~\ref{sec:strong_scaling}),
our hierarchical dynamic approach achieves speedups of up to $1.51$x, $1.83$x, and $1.91$x for different
scan algorithms, as presented in Figure~\ref{fig:strong_scan2},
while decreasing the overall energy consumption up to $2.23$x times (Section~\ref{sec:evaluation_energy}).

Our paper makes the following contributions:
\begin{itemize}
	\item A novel node-local, work-stealing prefix scan that exploits the hierarchy of parallel workers and memories to (1) decrease performance and energy costs of a distributed prefix scan and (2) exploit the additional levels of a shared-memory parallelization to construct an efficient load balancing step. To the best of our knowledge, this is the first scan algorithm designed for problems with unbalanced workloads.
	%We show that improving scalability of prefix scan on large number of cores.
	\item A scalable and efficient parallelization strategy for recursive image registration that enables analysis of temporal changes in long microscopy acquisitions. With our dynamic prefix scan, the performance of image registration is improved up to two times while decreasing energy costs  by over two times.
	%\item A scalable and efficient parallelization strategy for recursive image registration that enables the analysis of temporal changes in long microscopy acquisitions.
	%\item A novel example of a prefix scan problem where the sum operator is computationally intensive and exhibits huge variances in execution time.
  \item A novel example of a parallel scan problem, which performance challenges have not been addressed by research on parallel algorithms and MPI
    collectives, and a generic solution for expensive and unbalanced scan operators to fill this important gap in the quality of MPI collectives.
\end{itemize}

\section{Background and Motivation}
\label{sec:background}

Recent advances in transmission electron microscopy have allowed for a
more precise visualization of materials and physical processes, such as
metal oxidation, at nanometer resolution. Yet, many environmental factors
negatively affect the quality of microscopy images. A novel registration method\cite{Berkels201446} has been proposed
to mitigate these limitations by acquiring a series of low dose
microscopy frames and aligning each frame to the first frame
with an image registration procedure (Section~\ref{sec:image_registration}).
With this strategy, the increased amount of reliable information
extracted from noisy microscopy data is paid for by a computationally intensive and sequential process that
becomes a bottleneck of the analysis.
By phrasing the task of registering an image series as a special instance of the prefix scan (Section~\ref{sec:prefix_scan}), 
we can use the universal parallel pattern to propose parallelization strategies for this recursive computation.
We show that are no known prefix scans that can handle very well problems incorporating a high computation to communication ratio and an unpredictable and variable execution time (\ref{sec:related_work}).

\subsection{Prefix Scan}
\label{sec:prefix_scan}

%PPOPP
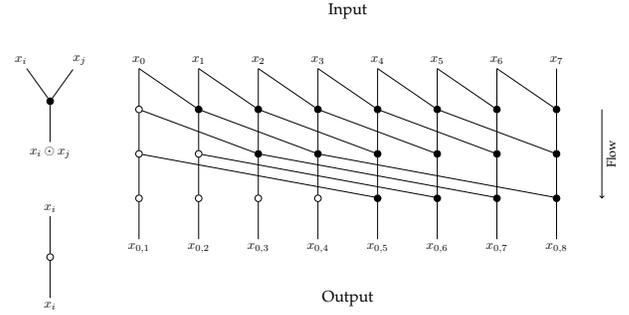
\begin{figure}[tb]
	\centering
	\resizebox{\dimexpr0.45\textwidth}{!}{%\documentclass{standalone}

%\usepackage{tikz,pgf, ifthen}
%\usetikzlibrary{backgrounds}
%\usetikzlibrary{calc, arrows,chains,positioning,scopes}

%\begin{document}

\def \n {7}

\begin{tikzpicture}[]%node distance  = 2 cm]
	every node/.style={anchor=north}
  %\useasboundingbox (-1,-1) rectangle (11,11); 
	\tikzset{VertexStyle/.style = {%draw, %shape          = circle,
                                 text           = black,
                                 %inner sep      = 2pt,
                                 %outer sep      = 0pt,
                                 %minimum size   = 29 pt
                                 }}
  \tikzset{PassiveStyle/.style = {draw, shape          = circle,
                         		text           = black,
                         		%inner sep      = 2pt,
                         		%outer sep      = 0pt,
                         		%minimum size   = 29 pt
                         		minimum size=5pt,
                         		inner sep=0pt
                         	}}
 \tikzset{ActiveStyle/.style = {draw, fill, shape          = circle,
                	  		text           = black,
                	  		%inner sep      = 2pt,
                	  		%outer sep      = 0pt,
                	  		%minimum size   = 29 pt
                	  		minimum size=5pt,
                	  		inner sep=0pt
                	  	}}
                                 
  \tikzset{FunctionStyle/.style = {
							  	  draw, shape = circle,
								  text = black
						  		  }}
  \tikzset{LabelStyle/.style =   {draw,
                                  fill           = yellow,
                                  text           = red}}
    % disable edge style for a moment
	\tikzset{EdgeStyle/.style   = {
		%very thick,
        %color=gray             		%double distance = 1pt
                     		%bend left
	}}
	\tikzset{EmptyStyle/.style = {draw=none
		}}                  
                           
    \node[VertexStyle](f0){$x_{0}$}; 
	\foreach \s [evaluate=\s as \si using \s-1] in {1,...,7}
	{
		\pgfmathsetmacro\sii{\s+1}
		\node[VertexStyle,right=of f\si](f\s){$x_{\s}$};
	}
	%\node[VertexStyle,right=of f1](f2){$\phi_{2, 3}$};    
	%\node[VertexStyle,right=of f2](fn1){$\phi_{N-2, N-1}$};
	%\node[VertexStyle,right=of f2](f3){$\phi_{3, 4}$};
	%\node[VertexStyle,right=of f3](f4){$\phi_{4, 5}$};
	%\node[VertexStyle,right=of f4](f5){$\phi_{5, 6}$};
	%\node[VertexStyle,right=of f5](f6){$\phi_{6, 7}$};
	%\node[VertexStyle,right=of f6](f7){$\phi_{7, 8}$};
     %\node[VertexStyle,below=of f1](f02){$\phi_{0,2}$};
     %\node[VertexStyle,below right=of f02](f03){$\phi_{0,3}$};
     %\node[VertexStyle,below right=of f03](f0N){$\phi_{0,N}$};
     %\node at ($(f2)!.5!(fn)$) {\ldots};
     %\node at ($(f2)!.5!(fn)$) {\ldots};
     
     %\node[VertexStyle,below=of f0](final1){$\phi_{0, 1}$};
     %\node[VertexStyle,right=of final1](final2){$\phi_{0, 2}$};
     %\node[VertexStyle,right=of final2](final3){$\phi_{0, 3}$};    
     %\node[VertexStyle,right=of f2](fn1){$\phi_{N-2, N-1}$};
     %\node[VertexStyle,right=of final3](finaln){$\phi_{0, N}$};
     %\node at ($(final3)!.5!(finaln)$) {\ldots};
     
     %\draw(f0) to node{}(final1);
     
	\node[PassiveStyle, below = of f0](B10){};
	\draw(f0) to node{}(B10);
	\foreach \s in {1,...,7}
	{
	   	\node[ActiveStyle, below = of f\s](B1\s){};
	   	\ifthenelse{\NOT\s=7}{
		  	\draw(f\s) to node{}(B1\s);
		}{	
			\draw[EdgeStyle](f\s) to node{}(B1\s);
		};
	}
	\foreach \s [evaluate=\s as \si using int(\s+1)] in {0,...,6}
	{
		\begin{scope}[on background layer]
		\ifthenelse{\NOT\s=0 \AND \NOT\s=4}{
			\draw(f\s.south) to node{}(B1\si);
		}{
			\draw[EdgeStyle](f\s.south) to node{}(B1\si);
		};
		\end{scope}
	}
    
    % second round of computation
 	\foreach \s in {0,...,1}
 	{
 		\node[PassiveStyle, below = of B1\s](B2\s){};
 		\draw(B1\s) to node{}(B2\s);
 	}
	\foreach \s in {2,...,7}
	{
		\node[ActiveStyle, below = of B1\s](B2\s){};
		\ifthenelse{\NOT\s=7}{
			\draw(B1\s) to node{}(B2\s);
		}{
			\draw[EdgeStyle](B1\s) to node{}(B2\s);
		}
	}
	\foreach \s [evaluate=\s as \si using int(\s+2)] in {0,...,5}
	{
		\begin{scope}[on background layer]
			\ifthenelse{\NOT\s=1 \AND \NOT\s=5}{
				\draw(B1\s) to node{}(B2\si);
			}{
				\draw[EdgeStyle](B1\s) to node{}(B2\si);
			}
		\end{scope}
	}
	
 	\foreach \s in {0,...,3}
 	{
 		\node[PassiveStyle, below = of B2\s](B3\s){};
 		\draw(B2\s) to node{}(B3\s);
 	}
	\foreach \s in {4,...,7}
	{
		\node[ActiveStyle, below = of B2\s](B3\s){};
		\ifthenelse{\NOT\s=7}{
			\draw(B2\s) to node{}(B3\s);
		}{
			\draw[EdgeStyle](B2\s) to node{}(B3\s);
		};
	}
	\foreach \s [evaluate=\s as \si using \s+4] in {0,...,3}
	{
		\begin{scope}[on background layer]
		\ifthenelse{\NOT\s=3}{
			\draw(B2\s) to node{}(B3\si);
		}{
			\draw[EdgeStyle](B2\s) to node{}(B3\si);
		}
		\end{scope}
	}
	
	\foreach \s in {0,...,7}
	{
		\pgfmathsetmacro\sii{\s+1}
		\node[VertexStyle,below=of B3\s](B4\s){$x_{0, \pgfmathprintnumber[int trunc]{\sii}}$};
		\ifthenelse{\NOT\s=7}{
			\draw(B3\s) to node{}(B4\s);
		}{
			\draw[EdgeStyle](B3\s) to node{}(B4\s);
		};
	}
	
   	\node[above=of $(f3)!0.5!(f4)$](){\large Input};
   	\node[below=of $(B43)!0.5!(B44)$](){\large Output};
   	
	\node[VertexStyle, left=of f0](Bj){$x_{j}$}; 
	\node[VertexStyle, left=of Bj](Bi){$x_{i}$};
	\node[ActiveStyle, below = of $(Bi)!0.5!(Bj)$](Bijapply){};
	\node[VertexStyle, below = of Bijapply](Bij){$x_{i} \odot x_{j}$};
	\draw(Bi) to node{}(Bijapply);
	\draw(Bj) to node{}(Bijapply);
	\draw(Bijapply) to node{}(Bij);
	
	\node[VertexStyle, below=of Bij](BiEM){$x_{i}$};
	\node[PassiveStyle, below = of BiEM](BiEMapply){};
	\node[VertexStyle, below = of BiEMapply](BjEM){$x_{i}$};
	\draw(BiEM) to node{}(BiEMapply);
	\draw(BiEMapply) to node{}(BjEM);
	
	\node[EmptyStyle, right=of B17](AxisUp){};
	\node[EmptyStyle, right=of B37](AxisDown){};
	\path[->,every node/.style={sloped,anchor=north,auto=false,rotate=180}] (AxisUp.center) edge  node {Flow} (AxisDown.center);
     %\node[anchor=west](final01) at (current page.west) {$\phi_{0,3}$};
     
     %\draw(f0) to node{}(f02);
     %\draw(f1) to node{}(f02);
     
     %\draw(f02) to node{}(f03);
     %\draw(f2) to node{}(f03);
     
     %\draw[dotted](f03) to node{}(f0N);
     %\draw(fn) to node{}(f0N);
     %\draw[EdgeStyle](f0) to node[LabelStyle]{$B$} (f1) ;
     %
     %\tikzset{EdgeStyle/.append style = {bend left = 60}}
     %\draw[EdgeStyle](f0) to node[LabelStyle]{$B$} (f2) ;
     %\draw[EdgeStyle](f0) to node[LabelStyle]{$B$} (fn1) ;
     %\draw[EdgeStyle](f0) to node[LabelStyle]{$B$} (fn) ;

\end{tikzpicture}

%\end{document}
}
	\caption{The dissemination prefix scan. Black dots represent an application of the operator while a white dot indicates a communication that does not involve any computation on the receiver. An optimal logarithmic depth is achieved by performing $N - 2^{i}$ operations in $i$-th iteration. For input data of size 8, 17 operator applications are necessary to obtain results in 3 iterations.}
	\label{fig:kogge_stone}
	%\vspace{-7mm}
\end{figure}
The importance and complexity of prefix scans make it one of the most
studied basic patterns in parallel computing. Numerous algorithms exist
that trade-off additional work and lower parallel
depth. The work--depth relation of prefix scans was an open research problem for many decades~\cite{Snir:1986:DTP:8088.8091, Zhu:2006:CZP:1142155.1142162}.
%
% PPOPP
The sketch of the dissemination prefix sum in Figure~\ref{fig:kogge_stone} depicts the main idea applied to parallelize the prefix scan: a decrease in depth is obtained by performing multiple computations on a single data element.
%
%An important distinction between different algorithms is the amount of work performed.
Depth--optimal algorithms cannot be zero--deficient~\cite{Zhu:2006:CZP:1142155.1142162}, i.e., an increase in work must be larger than the decrease in depth. Although depth minimization is the primary goal
when designing scalable algorithms, a huge work intensity usually implies an excessive communication.
Work--inefficient algorithms are more sensitive to deviations in execution time since they require more applications of the binary operator. Imbalanced operators will affect differently various scan algorithms due to differences in propagation of dependencies.
%The propagation of dependencies is different in various algorithms and it will differently affect the performance of applications with a non--constant time of an operator application.

%\review{
%}

%Internally, algorithms consist of several loop iterations, which are executed synchronously and each iteration depends on results from the previous one. Since many distributed algorithms are tuned for cases when the ratio of communication to \review{computation} is quite high, depth--optimal algorithms are often abandoned in favor of approaches that minimize the effects of network latency and congestion. Furthermore, work--inefficient algorithms are more sensitive to deviations in execution time since they require more applications of the binary operator. The propagation of dependencies is different in various algorithms and it will differently affect the performance of applications with a non--constant time of an operator application.

%\review{
A tree--based prefix scan is one of the classical parallel prefix scan strategies, as presented by Blelloch~\cite{Blelloch:1989:SPP:76108.76113} and Brent et al.~\cite{Brent:1980:CCB:800141.804666}. For both algorithms, the depth is bounded by a double traversal of a binary tree.
The dissemination prefix scan, also known as the recursive doubling~\cite{EGECIOGLU198995}, was presented by Kogge et al.~\cite{Kogge:1973:PAE:1638607.1639095} and Hillis et al.~\cite{Hillis:1986:DPA:7902.7903}.
The recursive family of prefix circuits presented by Ladner et al.~\cite{Ladner:1980:PPC:322217.322232} achieve an asymptotically smaller work overhead at optimal time but are rarely used in practice due to a less
favorable communication pattern.
For most of the well-known parallel prefix circuits, the depth is given as $C_{1} \log_{2}{N} + C_{2}$, where $C_{1}$ are $C_{2}$ are integer constants. The constant $C_{2}$ is non-zero for algorithms such as a tree-based inclusive scan presented by Brent et al., which has one layer less than the exclusive Blelloch scan. The Ladner-Fischer scan is designed with a constant $C_{2}$ that controls the depth--work balance.
%}

\begin{table}[tb]
	\centering
	%\setlengtbh{\dimexpr0.5\textwidth}
	%\setlength{\tabcolsep}{pt}
	%https://tex.stackexchange.com/questions/10535/how-to-force-a-table-into-page-width
	\begin{tabular*}{0.5\textwidth}{l @{\extracolsep{\fill}} cccc}
		\hline
		Name 			& Type & Depth					& Work \\ \hline \noalign{\smallskip}
		Sequential  	& I & $N - 1$					& $N-1$ \\
		Blelloch 		& E	& $2\log_{2}{N}$		& $2(N-1)$  \\
		Dissemination	& I	& $\log_{2}{N}$				& $N\log_{2}{N} - N + 1$ \\
		Ladner--Fischer	& I	& $\log_{2}{N}$				& $< 4N -5 $ \\
	\end{tabular*}
	\caption{Major \textit{I}-nclusive and \textit{E}-xclusive parallel prefix scan algorithms. The exact work for the Ladner--Fischer scan is given by a recursive equation in $N$.}
	
	% PPOPP
%	
%	\begin{tabular*}{0.5\textwidth}{l @{\extracolsep{\fill}} cccc}
%		\hline
%		Name 			& Type & Depth					& Work & Fan-out\\ \hline \noalign{\smallskip}
%		Sequential  	& I & $N - 1$					& $N-1$ & 1 \\
%		Blelloch 		& E	& $2\log_{2}{N}$		& $2(N-1)$ & 1 \\
%		Kogge--Stone	& I	& $\log_{2}{N}$				& $N\log_{2}{N} - N + 1$ & 1\\
%		Ladner--Fischer	& I	& $\log_{2}{N}$				& $< 4N -5 $ & $\dfrac{N}{2}$\\
%	\end{tabular*}
%	\caption{Major parallel prefix scan algorithms. Algorithms can be of \textit{I}-nclusive or \textit{E}-xclusive type. The exact work for the Ladner--Fischer scan is given by a recursive equation in $N$. Fan-out defines the maximal number of receivers for a single value.}
	\label{table:chap_prefix_sum_summary}
	
	%\vspace{-11mm}
\end{table}

%\review{
Table~\ref{table:chap_prefix_sum_summary} presents a comparison of the discussed parallel prefix scan algorithms.
Exclusive and inclusive variants are specified in the Message
Passing Interface standard as the collective operations
\texttt{MPI\_Exscan} and \texttt{MPI\_Scan}~\cite{mpi-3.0},
respectively. 
They are implemented using either simple algorithms that achieve the optimal
runtime of $\log_2 P$ rounds on $P$ processes
%at a cost of additional $\mathcal{O}(n\log n)$ work
or tree-based algorithms that optimize the communication latency~\cite{Sanders2006}. 
Most scan implementations are optimized for the common case that
\emph{communication time dominates computation time and that computation is
balanced}.

\subsection{Related Work}
\label{sec:related_work}

% change description from comparison with my own work to stating open research problems & challenges
% state that there has been no work on that topic that solves problems we work on

Standard strategies for a prefix scan when data size significantly exceeds the number of parallel workers have been frequently presented by other authors. Kruskal et. al~\cite{6312202} presented such algorithm on an EREW model~\cite{6312202}. It was later applied on a binary tree network of processors by Meijer~\cite{Meijer1987} and to solving a tridiagonal linear system on a hypercube architecture by Eǧecioǧlu et al.~\cite{EGECIOGLU198995}. There, the authors define the algorithm with a fixed choice of the dissemination as a global scan algorithm, whereas we present in Section~\ref{sec:distributed_prefix_scan} a generic distributed prefix sum and consider various scan algorithms in the global phase. 
Eǧecioǧlu et. al~\cite{EGECIOGLU199277} introduced a recursive algorithm for a distributed prefix scan, where it was found to have better efficiency than the previous approaches when the discrepancy between computation and communication cost is significant. This work was the first one to introduce more complex data distribution for prefix scan and to design a prefix scan strategy for computationally intensive operators. Data segments are distributed according to the number of arithmetical steps performed by each processor, an information that cannot be either estimated or predicted in problems such as image registration. Thus, designing a prefix scan for applications with an unknown load balance is an open problem.
Chatterjee et. al.~\cite{Chatterjee:1990:SPV:110382.110597} defined for the vectorization of prefix scan on CRAY-MP a strategy known as the \textit{reduce--then--scan}.
%The description of the algorithm is limited to an exclusive case, the global prefix scan is serialized and neither depth nor work are analyzed.
Although we use the same strategy as a basic for hierarchical and work--stealing scan, we extend it extensively with a dynamic accumulation of partial results and load balancing between neighboring threads.

%In the last decade, multiple papers were published on scan implementations for GPGPU architectures. The research is focused on optimizing scan operators for GPU execution model and its memory architecture, which makes their results not applicable to our case. There, the general algorithm used by us is known under the name "scan-then-propagate"~\cite{Sengupta:2007:SPG:1280094.1280110} and the alternative strategy is described as "reduce-then-scan"~\cite{Dotsenko:2008:FSA:1375527.1375559}.

%PPOPP
%\paragraph{Prefix scan in MPI}

Research on tuning \texttt{MPI\_Scan} and \texttt{MPI\_Exscan} collectives is focused on reducing the communication cost and improving bandwidth on memory--bound operators with computation being far cheaper than communication. Sanders et al.~\cite{Sanders2006} used pipelined binary trees and improved later the performance of the prefix scan in message-passing systems by exploiting a bidirectional communication~\cite{Sanders2007,SANDERS2009581}. The improvements are limited to prefix operators bounded by network latency, which is not the case for the image registration.

%\vspace{-3mm}

\subsubsection{Specific prefix scan operators} Although the prefix scan research has been dominated by optimizations dedicated to trivial operators, there has been few examples of prefix scans with computationally intensive operators. Maleki et al.~\cite{Maleki:2014:PDP:2692916.2555264} consider prefix scan solution to the linear-tropical
dynamic programming problem where the operation is a matrix-matrix multiplication. Gradl et al.~\cite{10.1007/11823285_78} presented a parallel prefix algorithm for accumulation of matrix multiplications in quantum control. Waldherr et al.~\cite{10.1007/978-3-642-13872-0_4} and Auckenthaler~\cite{AUCKENTHALER2010359} showed later that prefix scan parallelization of this operation is outperformed by a sequential prefix scan with parallel matrix multiplication operator. These applications of prefix scan resulted in neither tuning nor designing a scan algorithm for operators where computation time is significantly larger than communication.

In addition to performance improvements shown on the image registration problem, our work--stealing scan be applied to improve the efficiency of other imbalanced scans as well, and excellent examples of imbalanced operators are sparse linear algebra operations, found in the scan parallelization of neural network backpropagation with sparse matrix operations~\cite{DBLP:journals/corr/abs-1907-10134}. Prefix scans are essential for the automatic parallelization of loop-carried dependencies~\cite{10.1145/2259016.2259027}. While polyhedral techniques allow for approximating a balanced distribution of non-uniform loop nests, a dynamic work-stealing would improve the performance when static scheduling is not possible due to dynamic and data-dependent control-flow.

\begin{figure}[htb]
	\centering
	
	\resizebox*{0.9\width}{0.8\totalheight}{
		\subfloat[Frame 26]{%
			\includegraphics[width=\dimexpr0.22\textwidth-2\fboxrule,frame={\fboxrule}]{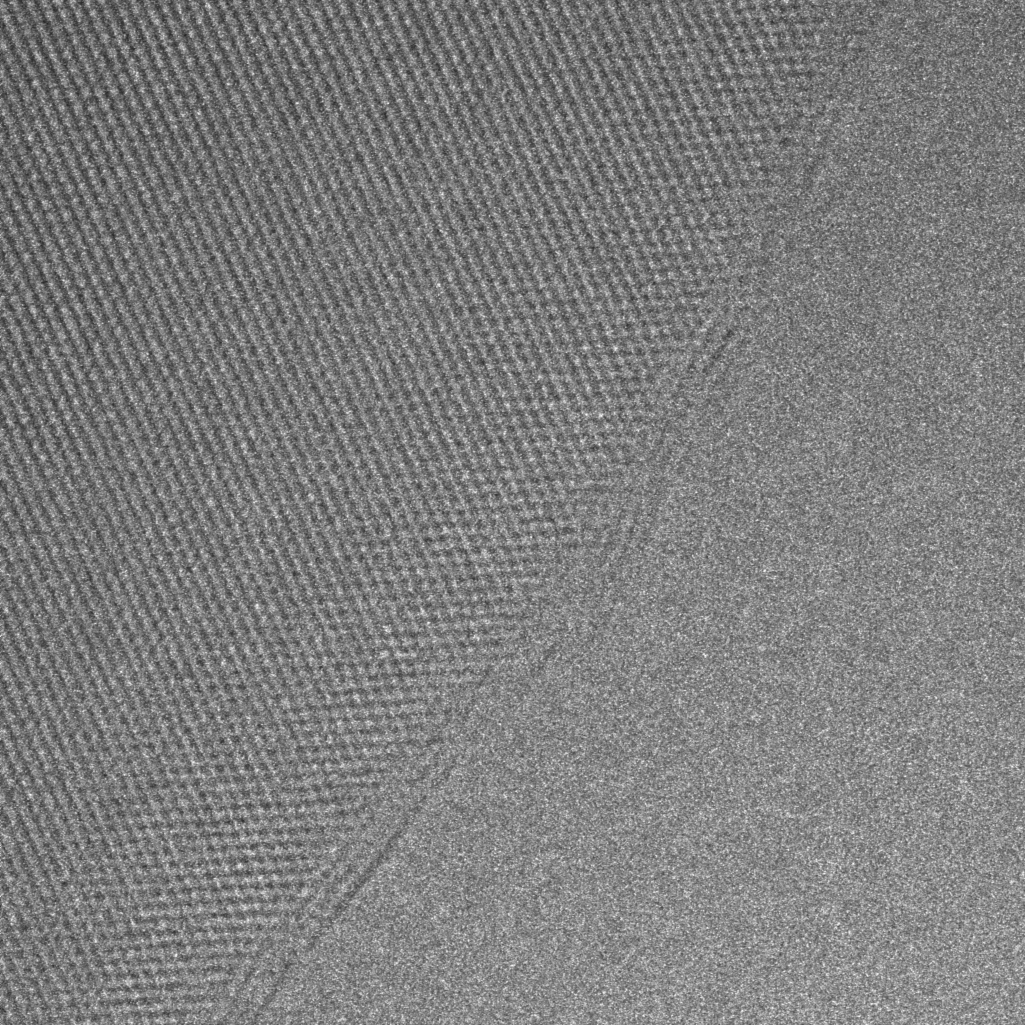}\label{fig:electron_data_ref}}
	}
	\hfill
	\resizebox*{0.9\width}{0.8\totalheight}{
		\subfloat[Deformed Frame 26, top left]{%
			\includegraphics[width=\dimexpr0.22\textwidth-2\fboxrule,frame={\fboxrule}]{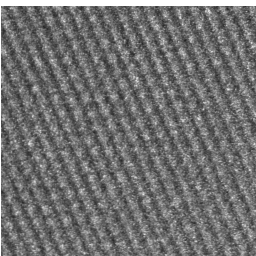}\label{fig:electron_data_def_crop}
		}\hfill
	}
	\caption{Frame 26 of the acquisition (left) and magnified after alignment to Frame 25 (right). The movement of the frame along vertical axis is visible as white stripe on the top. We observe a low variability between images acquired in a short timespan. TEM data courtesy of Sarah Haigh, University of Manchester.}  \label{fig:chapter_img_electron_data}
  \vspace{-5mm}
	%PPOPP
	%Frame 26 obtained in the twenty-first second of the first minute of the acquisition (left) and magnified after alignment to Frame 25 (right). The magnification makes the movement of the frame along the vertical axis visible as white stripe on the top. We observe a very low variability between two images acquired in a short timespan. TEM data courtesy of Sarah Haigh, University of Manchester.}  \label{fig:chapter_img_electron_data
\end{figure}

\subsection{Image Registration}
\label{sec:image_registration}

%The registration procedure
%estimates a transformation between each image pair but finding a
%transformation between any two frames requires a recursive accumulation
%of all subsequent transformations due to existence of periodic
%structures in the image.
%With this strategy, the increased amount of reliable information
%extracted from noisy electron microscopy data is paid for by an
%additional, computationally intensive and sequential process that
%becomes a bottleneck of the analysis. Each second of data acquisition
%might generate hundreds of frames and the registration method quickly
%becomes impractical for an acquisition running longer than a few dozen
%seconds. Therefore, by phrasing the task of registering an image series as a special instance of the prefix scan (Section~\ref{sec:prefix_image_registration}), we can use the universal parallel pattern to propose parallelization strategies for this recursive computation. We then show how vastly this setting differs from other known prefix sum problems, incorporating not only an unusually high computation to communication ratio but also an unpredictable and variable execution time.

We consider a series of two-dimensional, noisy atomic-scale electron microscopy images $f_{0}, f_{1}, \dots, f_{N}$ that are used instead of a single, high quality frame acquired with a high-dose electron beam.
Short-exposure image series are used since they allow to obtain a higher precision than a single image in the electron microscopy setting~\cite{YaBeDa14}. This replacement requires an aggregation of the information contained in the entire image series, usually done by averaging the images. However, the images cannot be averaged directly, since they are affected by environmental noise of the observed sample during acquisition. Considering that electron microscopy allows for a magnification by more than 10 million, even movements of the sample by just half of the width of an atom result in shifts of the observed images by several pixels. 
To mitigate the effects of sample drift, each frame is registered to the first image. Since the images are showing atomic grids, they have a high degree of self similarity in the form of (nearly) periodic structures, cf. \cref{fig:chapter_img_electron_data}. This periodicity makes the registration much more difficult: Given a pair of (nearly) periodic images without any prior information on their relative shift, registration can only determine the shift up to a multiple of the period of the images, which is not sufficient for the reconstruction. If the estimated shifts are off by a multiple of the period, unrelated positions will be averaged, which will blur and duplicate deviations, but this deviations are what is actually interesting for the applications, since they can significantly influence the material properties. The shift between non-consecutive images can be large, which prevents directly registering non-consecutive images in this setting.

The need for HPC arises when the temporal behavior of the observed sample needs to be studied. In this setting, a series can consist of hundred thousand or more high resolution images that are still subject to the problem of periodicity. As of now, such series are simply not analyzed as a whole but only manually selected subsets. Being able to analyze such series in their entirety has a large potential to lead to new insights in materials science that are otherwise inaccessible.

\subsubsection{Image Registration}
\label{par:image_registration}

We define the problem of image registration for two--dimensional images $\mathcal{R}, \mathcal{T} \in \mathcal{I}$, known as reference and template images, respectively.
\begin{definition}{\textbf{Image registration problem}}\label{def:image_reg} Given a distance measure $\mathcal{D}: \mathcal{I}\times \mathcal{I} \longrightarrow \mathbb{R}$ and two images $\mathcal{R}, \mathcal{T} \in \mathcal{I}$, find a transformation $\phi: \mathbb{R}^{2} \longrightarrow \mathbb{R}^{2}$ such that
	\begin{align*}
	\mathcal{D}(\mathcal{R}, \mathcal{T} \circ \phi)
	\end{align*}
	is minimized. $\phi(x) =  R(\alpha) \cdot x + G$ is a rigid transformation with angle $\alpha$, rotation matrix $R(\alpha) \in \mathbb{R}^{2 \times 2}$ and translation $G \in \mathbb{R}^{2}$.
	%PPOPPP
	%translaction vector
\end{definition}
Intuitively, we want to find $\phi$ such that the deformed template image is aligned to the reference: $\mathcal{T} \circ \phi \approx \mathcal{R}$. We use the image registration procedure proposed by Berkels et. al.~\cite{Berkels201446}. The approach defines a normalized cross-correlation functional as the distance measure and proposes a combination of a multilevel scheme with a gradient flow minimization process to solve the registration problem.
The objective functional is characterized by the presence of multiple local minima. The computed deformation may vary not only between different starting points for the minimization but also among various implementations of the same algorithm, resulting in unpredictable computation time. We refer to an implementation of this technique as the function \textbf{A}. It accepts two images with consecutive indices, $f_{i}$ and $f_{i+1}$, and estimates a deformation $\phi_{i, i+1}$ with the proposed algorithm.

%The multilevel process minimizes the likelihood of a gradient flow solver stopping at a local minimum by iteratively solving the problem on finer grids.
%A gradient flow solver is a generalization of the gradient descent, which permits computing the gradient with a respect to a generic scalar product.
%The objective functional is characterized by the presence of multiple local minima. The computed deformation may vary not only between different starting points for the minimization but also among various implementations of the same algorithm, resulting in unpredictable computation time. We refer to an implementation of this technique as the function \textbf{A}. It accepts two images with subsequent indices, $f_{i}$ and $f_{i+1}$, and an initial guess for the deformation $\phi_{0}$. The function applies the proposed algorithm to estimate a deformation $\phi_{i, i+1}$ as
%
%\begin{align*}
%\forall i \in \mathbb{N} \: \phi_{i, i + 1} &= \mathbf{A}(f_{i}, f_{i+1}, I_{\phi}),
%\end{align*}
%where $I_{\phi}$ denotes the identity transformation, which is a suitable initial guess for consecutive images.

\begin{figure}[htb]
	%\vspace{-6mm}
	\centering
	
	\resizebox*{0.9\width}{0.8\totalheight}{
		%\documentclass{standalone}
%
%\usepackage{tikz,pgf, ifthen}
%\usetikzlibrary{backgrounds}
%\usetikzlibrary{calc, arrows,chains,positioning,scopes}
%
%\begin{document}

\begin{tikzpicture}[]
	every node/.style={anchor=north}
	\tikzset{VertexStyle/.style = {%draw, %shape          = circle,
                                 text           = black,
                                 }}
	\tikzset{EdgeStyle/.style   = {
		%very thick,
        %color=gray             		%double distance = 1pt
        bend right=55,
        anchor=north,
	}}
    \node[VertexStyle](f and 0){$f_{0}$}; 
	\foreach \s [evaluate=\s as \si using \s-1] in {1,...,5}
	{
		\pgfmathsetmacro\sii{\s-1}
		\node[VertexStyle,right=of f and \si](f and \s){$f_{\s}$};
	}
	\foreach \j/\s in {0/1,1/2,2/3,3/4,4/5}
	%\foreach \j in {1,...,4}
	{
		\pgfmathsetmacro\sii{\s-1}
		%\path[EdgeStyle,->] (f and \j) {$\j\s$} edge[out=north,in=north]   (f and \s);
		\path[EdgeStyle,<-] (f and \s) edge node[below] {$\phi_{\j,\s}$}   (f and \j);
	}
	\foreach \s in {2,3,4,5}
	%\foreach \j in {1,...,4}
	{
		\pgfmathsetmacro\sii{(\s-2)*0.05}
		%\path[EdgeStyle,->] (f and \j) {$\j\s$} edge[out=north,in=north]   (f and \s);
		\path[EdgeStyle,<-] (f and \s) edge node[below , pos=0.53-\sii] {$\phi_{0,\s}$}   (f and 0);
	}
	%\node [] (f0) {B} edge[out=west,in=east] (f1);
\end{tikzpicture}

%\end{document}
	}
	%\vspace{-5mm}
	\caption{The image registration process for a series of 6 frames. For an image $f_{i}$, the result from its predecessor $\phi_{0,i-1}$ is combined with a neighboring deformation $\phi_{i-1, i}$.}
	% to register the image to the reference frame $f_{0}$.}
	\label{fig:series_registration}
	%\vspace{-5mm}
\end{figure}
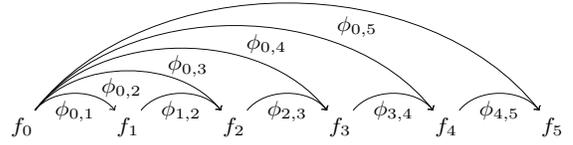

\subsubsection{Series Registration}

The alignment problem requires a dedicated approach when the images are (nearly) periodic and that is the case for electron micrographs. A correct registration of two frames is possible using the identity mapping as initial guess if the shift between them is smaller than half of the period. The validity of this assumption can be guaranteed only for neighboring frames $f_{i}$ and $f_{i+1}$. For the generic registration of $f_{0}$ and any frame $f_{i}$, this limitation can be bypassed by taking into account all neighboring frames in between where the procedure is deemed to be accurate. 

Given deformations $\phi_{0, 1}$ and $\phi_{1, 2}$ estimating $f_{1} \circ \phi_{0, 1} \approx f_{0}$ and $f_{2} \circ \phi_{1, 2} \approx f_{1}$, respectively, we can safely assume that the composition of deformations $\phi_{1,2} \circ \phi_{0, 1} $ is a decent initial guess to register $f_{0}$ and $f_{2}$, since
\begin{equation*}
\begin{split}
f_{2} \circ (\phi_{1,2} \circ \phi_{0, 1}) &= (f_{2} \circ \phi_{1, 2}) \circ \phi_{0, 1} \approx f_{1} \circ \phi_{0, 1} \approx f_{0}.
\end{split}
\end{equation*}
We approximate the deformation for two non--consecutive frames by using the composition of two deformations as an initial guess. We reuse the function $\mathbf{A}$ to define a new function $\mathbf{B}$
to handle non--consecutive indices %that operates on two deformations and their respective indices
$\phi_{i, k}  \: = \: \mathbf{B}(\phi_{i, j}, \; \phi_{j, k})$ and enable iterative registration up to $i$-th image, as shown in Figure~\ref{fig:series_registration}. 

\subsubsection{Associativity}
\label{sec:Associativity}

As we have already seen above, image registration is a non-convex 
optimization problem with multiple local minima. Thus, it may seem that 
the corresponding prefix scan operator is not associative. The special 
precautions we had to take for our specific setting with periodic 
structures are the key to get associativity in practice. This is due to 
our assumption that the shift between two consecutive images is smaller 
than half of the period and the way we construct initial guesses for the 
deformation for non-consecutive images, which should ensure that we 
start the minimization sufficiently close to the global minimum.
The registration process converges to correct results as long as deformations accumulate between adjacent images, ensuring that the shift between images is sufficiently small. Prefix scan preserves the guarantee, and thanks to the iterative optimization process, each operator application will converge to the best local solution even if changes in computation produce slightly different partial results.
The integrity of the data was verified with a manual inspection on small scale experiments, which included examples where various deformations provide equally suitable matches. A numerical comparison of cost function scores between sequential and parallel runs is not possible because the optimization process deals with a high level of noise in the input data and there, a different score does not necessarily indicate a worse or better match.

\section{Prefix Scan Image Processing}
\label{sec:prefix_image_registration}
%We formally define the problem as consisting of two steps - a preprocessing stage to generate neighbor transformations and the general registration for non--consecutive frames. The 
%\begin{definition}{\textbf{Image series registration}}\label{def:img_reg} Given a series of images $f_{0}, f_{1}, \dots, f_{n}$, perform a \textit{preprocessing} step to register each pair of frames
%	\begin{align*}
%	\phi_{i, i + 1} &= \mathbf{A}(f_{i}, f_{i+1}, I_{\phi})
%	\end{align*}
%	and the \textit{series registration} step to align each image $f_{i}$ to the reference $f_{0}$.
%\end{definition}
%Although the first step of registration is embarrassingly parallel due to lack of dependencies, the second phase requires a more sophisticated approach. In this paper, we interpret this stage as a prefix sum problem to parallelize it.

%\subsection{Prefix Sum Image Registration}
%\label{sec:prefix_image_registration}

% Copik: not sure where to put this, perhaps somewhere in next section
%\paragraph{Prefix Sum Formulation}

\begin{figure*}[htb]
	\centering
	\resizebox*{0.9\width}{0.8\totalheight}{
		\subfloat[Execution times of the image registration operator.]{%
			\includegraphics[width=\dimexpr0.49\textwidth]{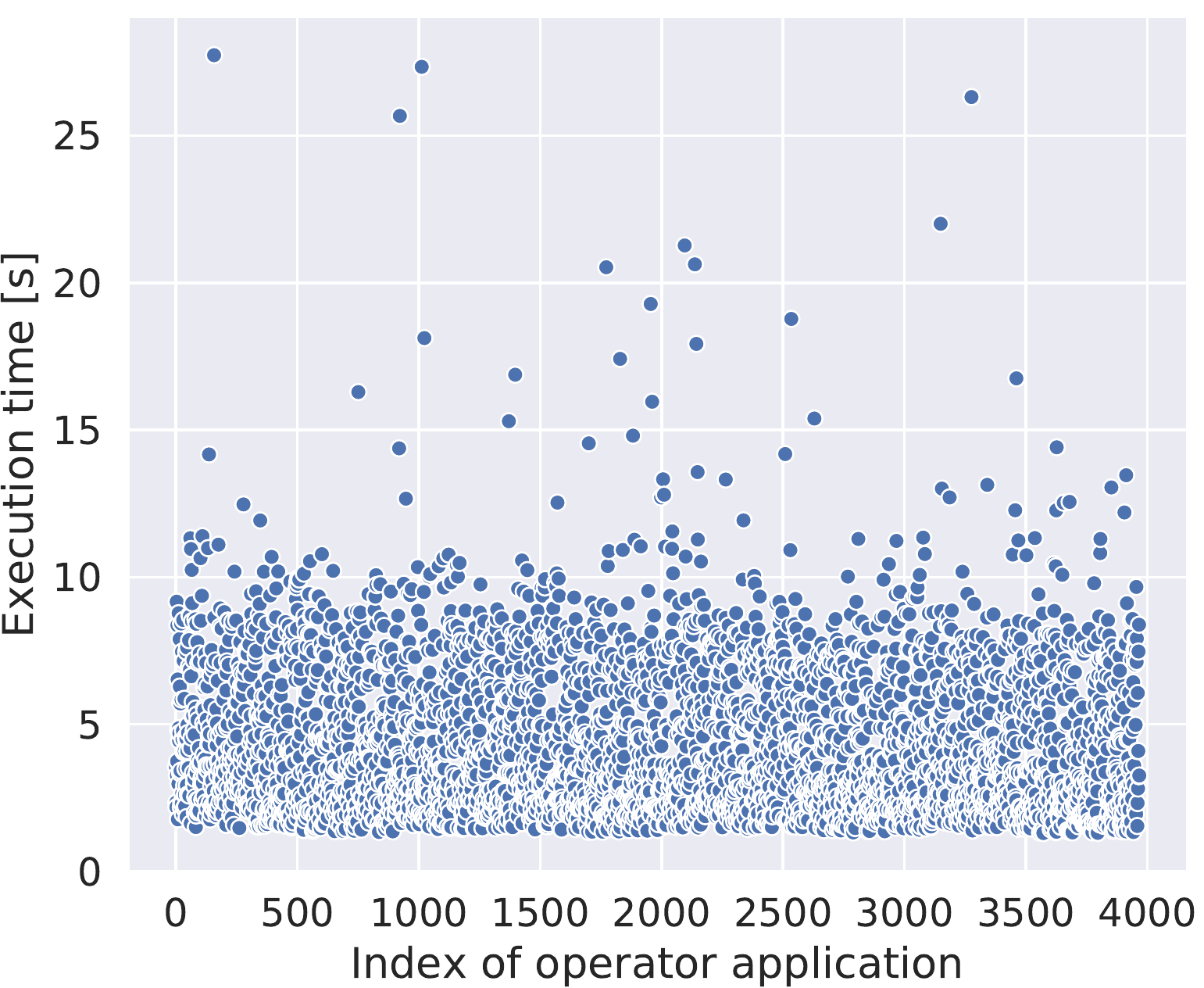}\label{fig:registration_times}}
		}
	%\hfill
	\resizebox*{0.9\width}{0.8\totalheight}{
		\subfloat[Relative comparison of maximum and mean segment execution time.
		%PPOPP
		%Relative comparison of maximum and mean segment execution time depending on the data segment size.
		]{%
			\includegraphics[width=\dimexpr0.49\textwidth]{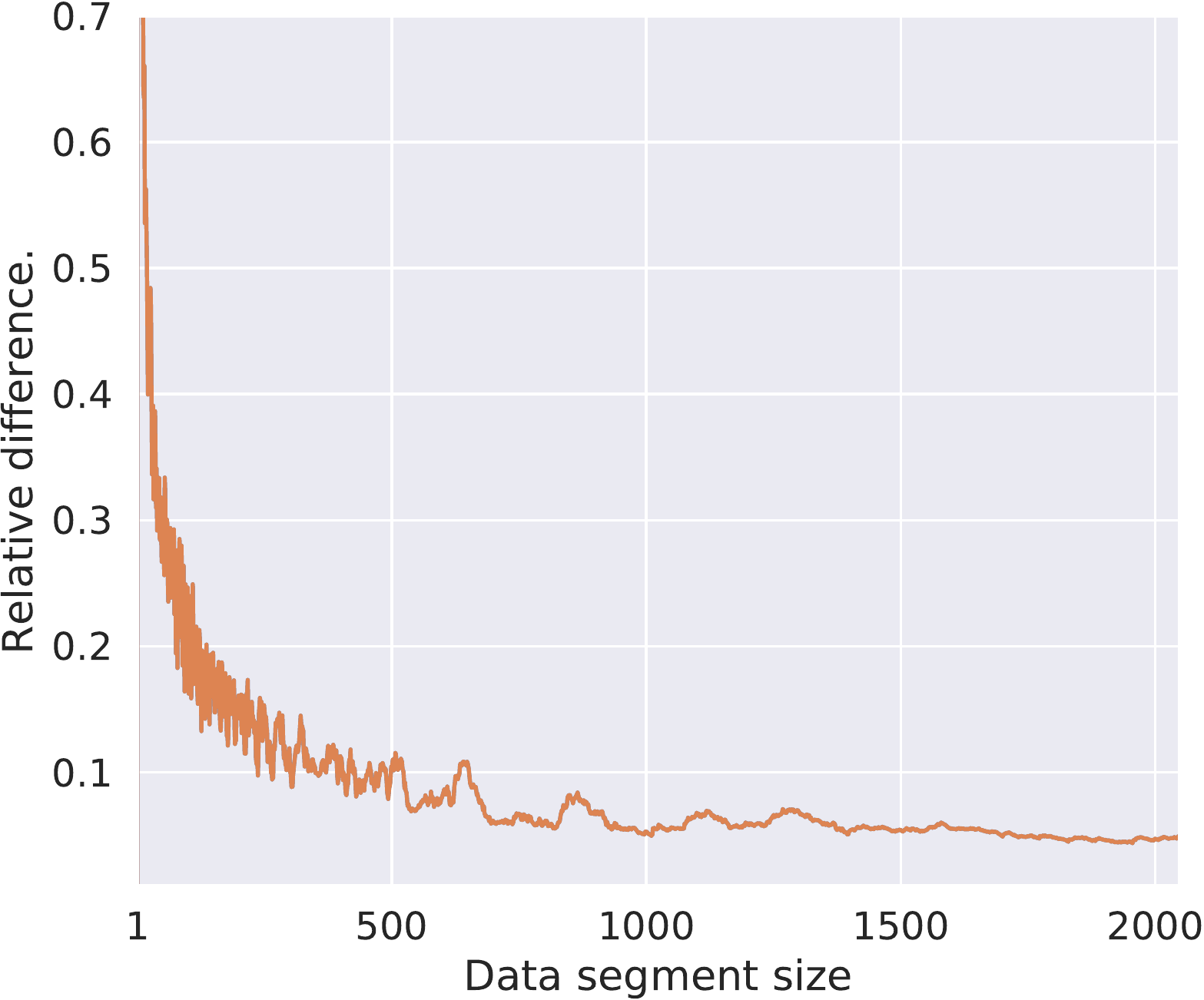} \label{fig:imbalance}}
	}
	\caption{
		The computationally intensive image registration operator: (a) execution times $t_{1}, \dots , t_{N}$ of the operator in the first local phase of the prefix scan and (b) load imbalance for a static distribution with segment size $S = \frac{N}{P}$, where
		%	for each segment $s$ with execution times $t_{b_{s}}, \cdots, t_{e_{s}}$ and ($T_{s} = \sum_{i=b_{s}}^{e_{s}} t_{i}$), 
		we estimate a relative difference between the mean ($\mu = \frac{S}{N} \sum T_{s}$) and maximum execution time ($\max_{s} T_{s}$) across all segments. The measurements were obtained on an Intel E5-2690 v3 CPU with 2.60 GHz base frequency.
	}
	% PPOPP	
	%The computationally intensive image registration operator. The presented data includes (a) execution times $t_{1}, \dots , t_{N}$ of the image registration operator in the first local phase of the prefix scan and (b) an analysis of load imbalance for a static distribution with segment size $S = \frac{N}{P}$, where for each segment $s$ with execution times $t_{b_{s}}, \cdots, t_{e_{s}}$ and ($T_{s} = \sum_{i=b_{s}}^{e_{s}} t_{i}$), we estimate a relative difference between the mean ($\mu = \frac{S}{N} \sum T_{s}$) and maximum execution time ($\max_{s} T_{s}$) across all segments. The measurements were obtained from registration of 4096 images on an Intel E5-2690 v3 CPU with 2.60 GHz base frequency. }

\end{figure*}

In the formulation of the series registration problem above, the recursive nature is immediately seen: for any image $f_{i}$, the task of aligning to $f_{0}$ requires solving the registration problem for $f_{0}$ and $f_{i-1}$ first. Each final deformation $\phi_{0, i}$ can be obtained by consecutively applying the registration algorithm to neighboring deformations $\phi_{0, 1}, \phi_{1, 2}, \dots, \phi_{i - 1, i}$. The accumulation of partial solutions can be represented as a prefix scan with the operator $\odot_{B}$ defined as follows
\begin{equation*}
	\phi_{i, j} \odot_{B} \phi_{j, k} = \mathbf{B}(\phi_{i, j}, \phi_{j, k}) \quad
	\phi_{0, j} = \phi_{0, 1} \odot_{B} \dots \odot_{B} \phi_{j-1,j}
\end{equation*}
% PPOPPP
%\begin{align*}
%\phi_{i, j} \odot_{B} \phi_{j, k} \: &= \: \mathbf{B}(\phi_{i, j}, \phi_{j, k}) \\
%\phi_{0, j} \: &= \: \phi_{0, 1} \odot_{B} \phi_{1, 2} \odot_{B} \dots \odot_{B} \phi_{i-2, i-1}
%\end{align*}
Clearly, this operator is not commutative and it does not have an inverse.
The prefix scan operator inherits all properties from the iterative registration method.
%PPOPP
%The registration function is encapsulated within the prefix scan operator, which in turn inherits all properties from the iterative registration method.
Therefore, the sum operator contains two distinct features that set it apart from most of the other problems analyzed in the context of prefix scan parallelization: (1) an unusually large ratio of computation to communication cost and (2) unpredictable execution time causing load imbalance issues.

\subsection{Computation Cost}

The simplest case of a prefix scan operator found in the literature, which happens to be the one most frequently evaluated, is integer addition.
%, which should not take more than few CPU cycles on modern processors.
More complex examples still involve relatively cheap operations, such as polynomial evaluation and  addition of summed area tables with multiple integer and floating--point multiplications. As a result, parallel prefix scan algorithms tend to be optimized for memory--bound operators with a low execution time.
Image registration does not fit into this category, as it can be seen in Figure~\ref{fig:registration_times}. A single operator application usually takes up to 10 seconds, with noticeable outliers going for up to 30 seconds. The resulting deformation stores only three floating--point values and the cost of sharing such data is dominated by the latency. The computation time is much larger than latency introduced by network communication and this discrepancy will not change even with significant serial optimizations of the operator.

\subsection{Load Imbalance}
\label{sec:load_imbalance}

Another particular feature of prefix scan operators that is commonly seen in the literature, is a deterministic execution time that does not change between applications. In contrast to operations with a predictable and constant runtime, here the actual computation cost is not only unpredictable but highly variant. Due to the iterative nature of the registration algorithm, we can not foresee for a given input data how many iterations are necessary to reach a stopping criterion. The time measurements presented in Figure~\ref{fig:registration_times} show that significant outliers do not form any regular distribution and estimation of an efficient distribution is not possible.
%A knowledge about such a distribution would be necessary to estimate load imbalance and select a more efficient data distribution before computation.
For the same dataset, we studied the load imbalance of a static data distribution to learn how the distributed run might be affected when the increase in computing resources leads to smaller data chunks available to each rank. We look at the difference between mean and maximum execution time across data segments. Intuitively, if the imbalance of computational effort between segments is large, then the larger is the difference between mean completion time and the slowest worker.
The results in Figure~\ref{fig:imbalance} show how the increase in execution time raises from roughly 5\% for large data segments to over 20\% when each segment contains less than 100 deformations. 
These results indicate how speedups of our parallel image registration are going to change when we scale the problem to the point where only a few dozens of images are available per MPI rank. The performance is going to decrease not only because of the raising cost of a global scan but also due to increasing influence of load imbalance. Since we want our configuration to be located on the part of the plot with a low imbalance factor, we have to choose a sufficiently large segment size. To that end, we present a hierarchical decomposition of prefix scan in Section~\ref{sec:hierarchical_prefix_sum} to group parallel workers and increasing the segment size on certain levels of hierarchy.

\section{Distributed Scan Strategies}
\label{sec:distributed_prefix_scan}
Prefix scan has been successfully applied in distributed and accelerated computations and many of
these attempts rediscovered the same standard strategies, \textit{scan--then--map} and \textit{reduce--then--scan},
when data size significantly exceeds the number of workers (Section~\ref{sec:distributed_prefix_scan2}).
We show how these strategies can be extended to a hierarchy of parallel workers (Section~\ref{sec:hierarchical_prefix_sum}),
such as the one used in a hybrid computation with multiple threads per each distributed worker,
with neither a loss of generality nor an increase of algorithm depth.
By defining a novel \textit{dynamic hierarchical prefix scan} (Section ~\ref{sec:dynamic_prefix_sum}),
we exploit the additional level of a shared-memory parallelization to provide a load balancing step,
%We show how \textit{reduce--then--scan} strategy can be used to change dynamically the work distribution
%among threads and improve performance of scan operators with an unbalanced and unpredictable execution
%time, such as the one introduced in the previous section.
improving performance of scan operators with an unbalanced and unpredictable execution
time, such as the one introduced in the previous section.

%\vspace{-4mm}
\begin{figure}[htb]
	\vspace{-5mm}
	\centering
	
	\resizebox*{0.9\width}{0.8\totalheight}{
    \subfloat[\textit{scan--then--map} prefix scan.]{%
      \includegraphics[width=\dimexpr0.49\textwidth]{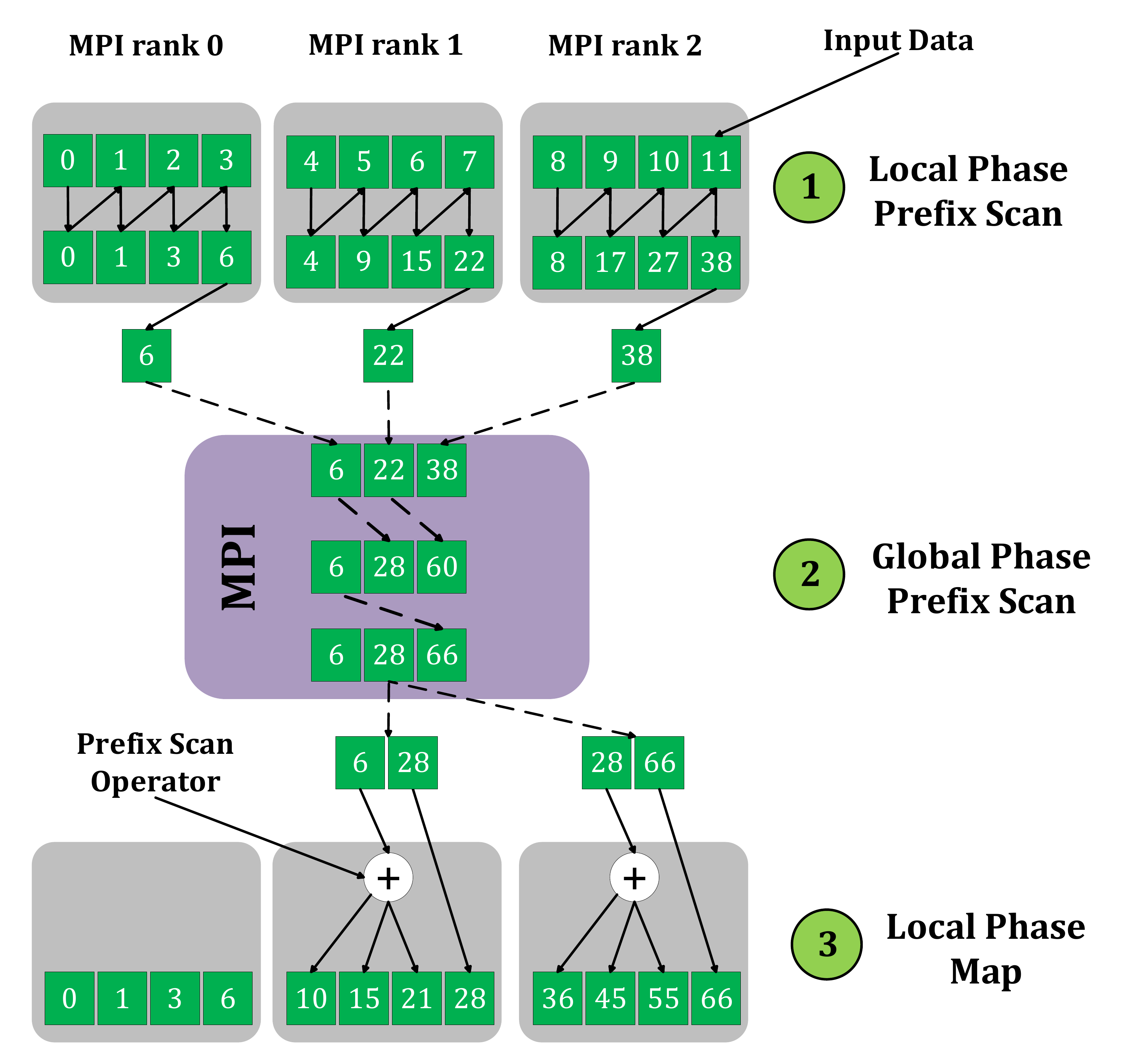} \label{fig:scan_then_map}%\hfill
    }
	}
	\resizebox*{0.9\width}{0.8\totalheight}{
    \subfloat[\textit{reduce--then--scan} prefix scan.]{%
      \includegraphics[width=\dimexpr0.49\textwidth]{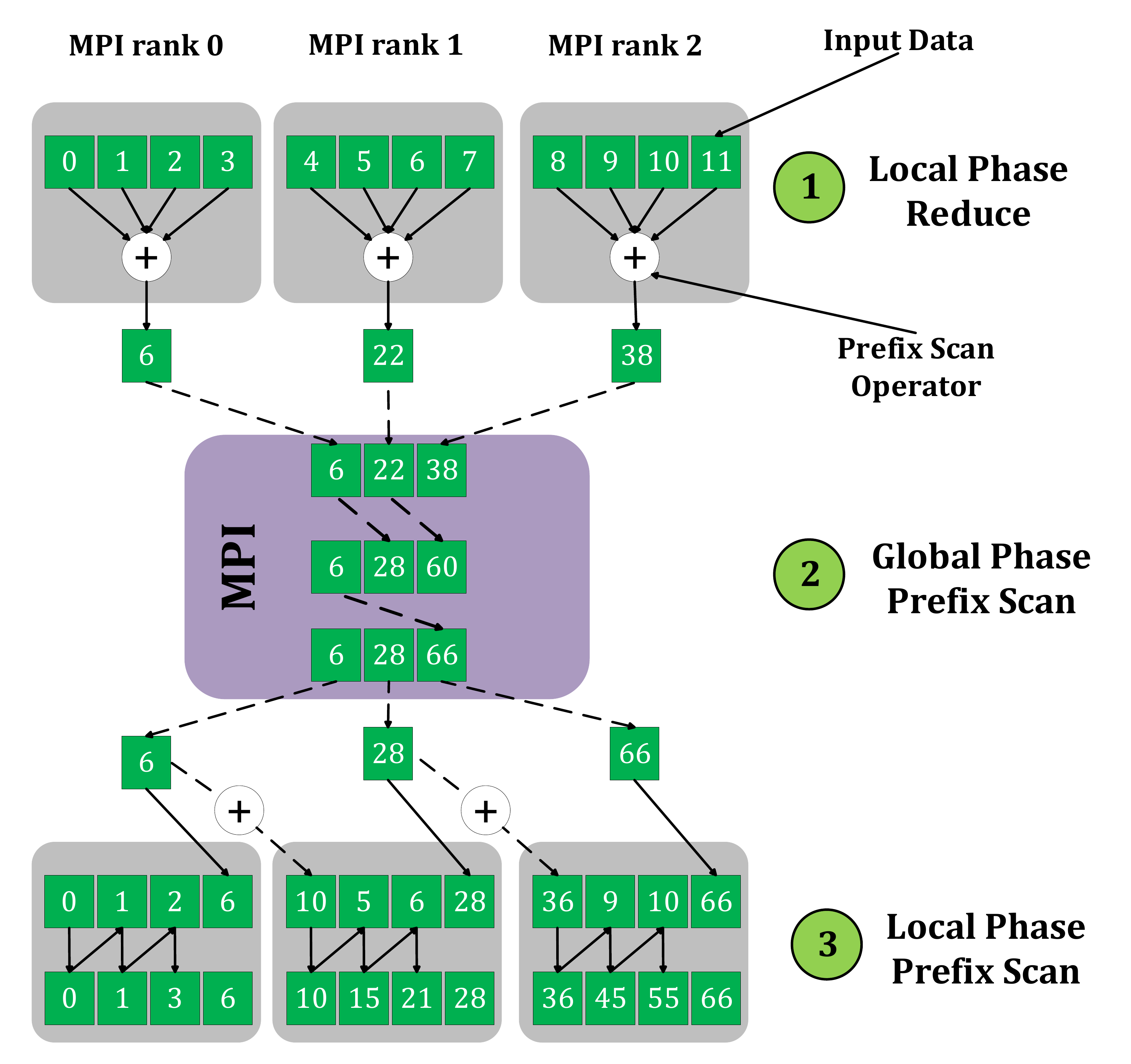} \label{fig:reduce_then_scan}%\hfill
    }
	}
	%\vspace{-4mm}
	%\caption{An example of a \textit{reduce--then--scan} distributed prefix scan for integer addition on three MPI ranks. The inclusive result from the global scan is used as a last local result and an exclusive result is received from the left neighbor. Operations local to MPI rank are shown with~\protect\includegraphics[scale=0.25,trim=18 15 10 18,clip]{images/full-arrow.pdf} whereas ~\protect\includegraphics[scale=0.25,trim=18 15 10 18,clip]{images/dashed-arrow.pdf} indicates global communication.}  \label{fig:distr_prefix_sum}
	\caption{Examples of distributed prefix scan strategies for integer addition on three MPI ranks. Operations local to MPI rank are shown with~\protect\includegraphics[scale=0.25,trim=18 15 10 18,clip]{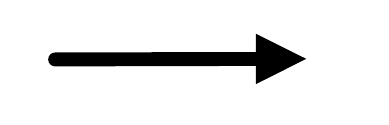} whereas ~\protect\includegraphics[scale=0.25,trim=18 15 10 18,clip]{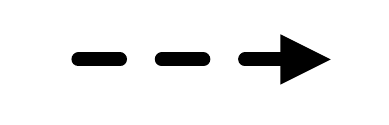} indicates global communication.}  \label{fig:distr_prefix_sum}	
	\vspace{-5mm}
\end{figure}

\subsection{Distributed Prefix Scan}
\label{sec:distributed_prefix_scan2}

The classical parallel prefix scan algorithms were designed to minimize the depth when the number of processing elements is equal to the number of data elements. Although such case is common in circuit design, it is not well-suited as a general solution since the length of input sequence $x_{0}, x_{1}, \dots, x_{N-1}$ is usually larger than then the number of workers $P$. For simplicity, we limit the analysis to the case of even data distribution, and each worker is assigned $K = \frac{N}{P}$ input elements with boundary indices $l_{I}$ and $r_{I}$.

The main strategy for a distributed scan follows a principle of splitting the work to local-global-local sequence of computations, as presented in Figure~\ref{fig:distr_prefix_sum}. First, each rank is assigned a data segment to process independently in the local phase ~\protect\includegraphics[scale=0.13,trim=18 18 10 18,clip]{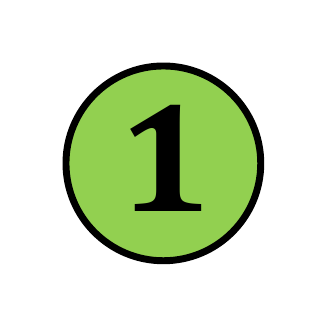}, computing a sum of all elements in the local segment $x_{l_{I}, r_{I}}$. The result is passed to a global prefix scan of size $P$, computing an accumulated result $x_{0, r_{I}}$ on each rank~\protect\includegraphics[scale=0.13,trim=18 18 10 18,clip]{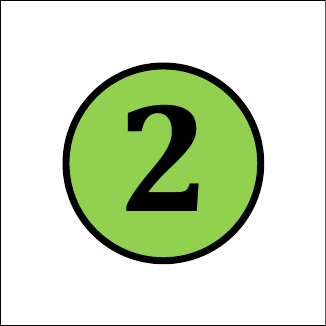}. The computation is finalized with an update of local data segments with an accumulated sum of $x_{0, r_{I-1}}$~\protect\includegraphics[scale=0.13,trim=18 18 10 18,clip]{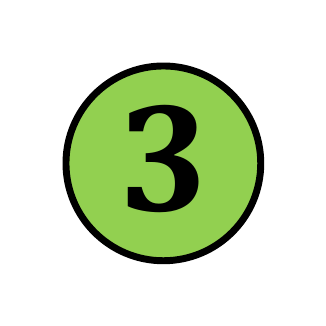}.
As global phase~\protect\includegraphics[scale=0.13,trim=18 18 10 18,clip]{images/dot_2.pdf}, one can
use any distributed scan implementation, such as \code{MPI\_Scan}.
The local phases~\protect\includegraphics[scale=0.13,trim=18 18 10 18,clip]{images/dot_1.pdf} and~\protect\includegraphics[scale=0.13,trim=18 18 10 18,clip]{images/dot_3.pdf} can be defined in two ways, either as a scan that updates local data and requires only adding global result in the end or as a reduction that leaves the segment intact and finishes the computation with a prefix scan.
The \textit{scan--then--map} procedure is usually preferred over the \textit{reduce--then--scan} approach since the former exhibits a slightly lower depth and decreased workload due to the first parallel worker inactivity in last phase. We use the former approach for evaluation of a standard, distributed prefix scan. %Both approaches differ substantially in flexibility.
%We use it for evaluation of a standard distributed prefix scan without neither hierarchy nor load balancing. 
%
%While in the scan--based algorithm the work distribution has to be determined before starting computations, replacing the first phase with a reduction allows to dynamically determine workload per thread during accumulation of partial results.
While in the \textit{scan--then--map} algorithm the work distribution has to be defined before execution, a dynamic determination of workload per thread is possible when the first phase is a reduction.
Such property is desirable for imbalanced computations to allow work-stealing and decrease disproportions in workload. In next two paragraphs we discuss advantages and disadvantages of each approach in detail.

\subsubsection{Scan--then--map}

In this approach, presented in Figure~\ref{fig:scan_then_map}, a \textit{scan} is computed initially in the first phase $LP_{1}$, creating a new sequence of partial results that requires only an \textit{application} of global scan result in the last phase. Depth and work of the first phase are straightforward:
%\vspace{-2mm}
\begin{equation*}	
	%S_{LS1}(N, P) &= S_{S}(\left \lceil \frac{N}{P} \right \rceil) = \left \lceil \frac{N}{P} \right \rceil - 1
	D_{LP1}(N, P) = \dfrac{N}{P} - 1 \quad W_{LP1}(N, P) = P \cdot \left( \dfrac{N}{P} -1 \right) = N - P
\end{equation*}
%PPOPP
%\begin{align*}	
%%S_{LS1}(N, P) &= S_{S}(\left \lceil \frac{N}{P} \right \rceil) = \left \lceil \frac{N}{P} \right \rceil - 1
%D_{LP1}(N, P) &= D_{Scan}\left( \dfrac{N}{P} \right) = \dfrac{N}{P} - 1 \\
%W_{LP1}(N, P) &= P \cdot D_{Scan}\left( \dfrac{N}{P} \right) = N - P
%\end{align*}

The last element computed by the local scan $x_{l_{I}, r_{I}}$ is the sum needed for a global scan. In the second local phase, each local result $x_{l_{I}, j}$ is combined with the exclusive value $x_{0, l_{I} - 1}$. An exception is the data segment assigned to worker $0$ which is already finished. This requires exactly $K$ applications of the operator. However, since the prefix scan is inclusive, the last value $x_{0, r_{I}}$ is already computed in the global phase which saves one application of the operator:
%\vspace{-2mm}
\begin{equation*}
D_{LP2}(N, P) = \dfrac{N}{P} - 1  \quad
W_{LP2}(N, P) = (P - 1) \cdot \left( \dfrac{N}{P} - 1 \right)
\end{equation*}
% PPOPP
%\begin{align*}
%\begin{split}
%D_{LP2}(N, P) &= \dfrac{N}{P} - 1 = D_{LP1}(N, P) \\
%W_{LP2}(N, P) &= (P - 1) \cdot \left( \dfrac{N}{P} - 1 \right) = N - P - \dfrac{N}{P} + 1
%\end{split}
%\end{align*}

The depth and work of the algorithm are given as follows:
\begin{align}
D_{DS}(N, P) &= D_{LP1}(N, P) + D_{GS}(P) + D_{LP2}(N,P) \nonumber \\
             &= 2 \cdot \frac{N}{P} - 2 + D_{GS}(N, P)
\end{align}
\begin{align}
W_{DS}(N, P) &= W_{LP1}(N, P) + W_{GS}(P) + \cdot W_{LP2}(N,P) \nonumber \\
             &= 2 \cdot N - 2 \cdot P - \frac{N}{P} + 1 + W_{GS}(N, P)
\end{align}
The analysis of critical path is possible for an even distribution of data. Otherwise, critical paths of local phases might be provided by different workers and simple summation would yield an incorrect result.
% PPOPP
%We must remark that this analysis of critical path is possible only in the condition of an even distribution of data. Otherwise, a simple summation of the depth for two local phases might yield an incorrect result if critical paths for those phases are provided by different workers.
%An example of such a situation may be a prefix scan where $N \bmod P = 1$. There, the span of the first phase is given by the first worker who has one more data element than other processes, but it is inactive in the second local phase.
%
The last phase can be parallel and balanced since each element is updated independently. Yet, there's no possibility to decrease load imbalance before global phase. 
% PPOPP
%A benefit of this strategy is the parallelizable last phase where each element can be updated independently. However, 
%This helps to resolve potential issues with load balance by freely distributing operator applications across workers.

\subsubsection{Reduce--then--scan}

As depicted in Figure~\ref{fig:reduce_then_scan}, each worker computes sequentially a \textit{reduction} in the first phase, leaving local data elements untouched until the \textit{scan} in the second local phase. There, the global result $x_{0, l_{I} - 1}$ is added to the first local element $x_{l_{I}}$ and the scan updates each value with $x_{0, l_{I} - 1}$. For the first phase, workload and depth does not change since the very first element $x_{l_{I}}$ can be used as an initial value of the sum. There's a difference in last phase, however. Although we can still use the trick with inclusive result, one needs first to apply the global result to the first element and the first worker is no longer idle in that phase.

\begin{equation*}
		D_{LP2}(N, P) = \frac{N}{P} \quad 
		W_{LP2}(N, P) = P \cdot \frac{N}{P} = N
\end{equation*}

% PPOPP review
%\begin{align*}
%\begin{split}
%D_{LP2}(N, P) &= \frac{N}{P} \\
%W_{LP2}(N, P) &= P \cdot \frac{N}{P} = N
%\end{split}
%\end{align*}

The depth and work of the algorithm are given as follows:
%\begin{align*}
%D_{DS}(N, P) &= D_{LP1}(N, P) + D_{GS}(P) + D_{LP2}(N,P) \\ &= 2 \cdot \frac{N}{P} - 1 + D_{GS}(N, P) \\
%W_{DS}(N, P) &= W_{LP1}(N, P) + W_{GS}(P) + W_{LP2}(N,P) \\ &= 2 \cdot N - P  + W_{GS}(N, P)
%\end{align*}
%\vspace{-4mm}
\begin{align}
\label{eq:reduce_then_scan}
D_{DS}(N, P) &= 2 \cdot \frac{N}{P} - 1 + D_{GS}(N, P) \\
W_{DS}(N, P) &= 2 \cdot N - P  + W_{GS}(N, P)
\end{align}
The increase in performed work is proportional to the number of workers. Contrary to the other approach, here the first phase allows for further parallelization due to less strict nature of reduction. The strictly sequential last phase is a minor disadvantage.
%PPOPP
%The increase in performed work is exactly $P + \dfrac{N}{P} - 1$ applications of operator. Thus, it is proportional to the number of workers. Contrary to the other approach, here the first phase allows for further parallelization due to less strict nature of reduction. However, the last phase is entirely sequential and work cannot be easily redistributed in case of a load imbalance.

\subsection{Hierarchical Prefix Scan}
\label{sec:hierarchical_prefix_sum}

We now present a novel strategy for a distributed scan that includes a hierarchical distribution of work and data. We show that such redistribution can be performed with a constant increase in the algorithmic depth in the worst case.
Even though we do not achieve reduction in depth, the hierarchization decreases the number of ranks participating in the global scan. This change reduces negative performance effects of an unbalanced global scan on many ranks and decreases the pressure and dependence on network communication by performing more computation intra-node. Moreover, applying the hierarchical scan to distributed computation introduces a lower hierarchy layer with shared--memory environment that allows for an efficient implementation of work stealing, as discussed in Section~\ref{sec:dynamic_prefix_sum}.
Although we consider here the most common case of a hybrid MPI application with local threads assigned to each rank, the general principle extends to an arbitrary number of levels.

For a multithreaded implementation, we assume that an allocation of $P$ MPI ranks is replaced with $P'$ ranks and $T$ threads such that $P' \cdot T = P$. 

\begin{enumerate}
	
	\item Local Phase on $P' \cdot T$ workers.
	
	For both scan and reduce, there is no change in either depth or work performed since each segment of size $\frac{N}{P}$ is replaced with a new one of length $\frac{N}{P' \cdot T}$.
	
	\item Local Scan on $T$ local segments.
	
	We assume that internally each rank uses the same parallel prefix scan algorithm as in the global scan, with $D' = C_{1} \log_{2}{T} + C_{2}$ and work $W' = f(T)$.
	
	\item Global Scan on $P'$ ranks.
	
	In each prefix scan iteration, the result received from other rank is applied by $T$ threads to $T$ scan results corresponding to inclusive prefix scan over all segments. Only the last result is used for communication.
	
	\item Second Local Phase on $P' \cdot T$ workers.
	
	The entire computation proceeds without major changes since each thread owns the scan result continuously updated in the global phase.

\end{enumerate}

We observe that the composition of local and global parallel scans does not change the asymptotic performance since $C_{1}\log_{2}{T} + C_{1} \log_{2}{P'} =  C_{1} \log_{2}{P}$ and only change is visible in constants $C_{2}$. Nevertheless, this increase in depth does not apply to depth--optimal scans, where $C_{2} = 0$, which are of special interest.
%PPOPP - repeated conclusion
%Although the algorithmic depth does not change significantly, we can expect an improved performance due to a lower dependence on inter-node communication and effects of network congestion.
%Therefore, the segment size is increased by a factor equal to a number of local workers and Figure~\ref{fig:imbalance} shows how significant improvements can be made when moving away from a small data segment size.

% not tested in code (not enough time)
%\paragraph{Optimizations}
%
%Furthermore, the hierarchical algorithm adds the ability to remove the last synchronization step of exchanging the inclusive results between neighbors. In the hierarchical case we have a strict distinction between $T - 1$ local results used internally and the global one that is communicated with other ranks. This allows us to use $T - 1$ threads to compute exclusive results in each iteration of a global scan and move to the last local phase immediately after receiving and processing the last message. The first segment still has to receive the exclusive result from the neighbor but this cna 

\begin{figure}[htb]
	%\vspace{-3mm}
	\centering
	\includegraphics[width=\dimexpr0.49\textwidth]{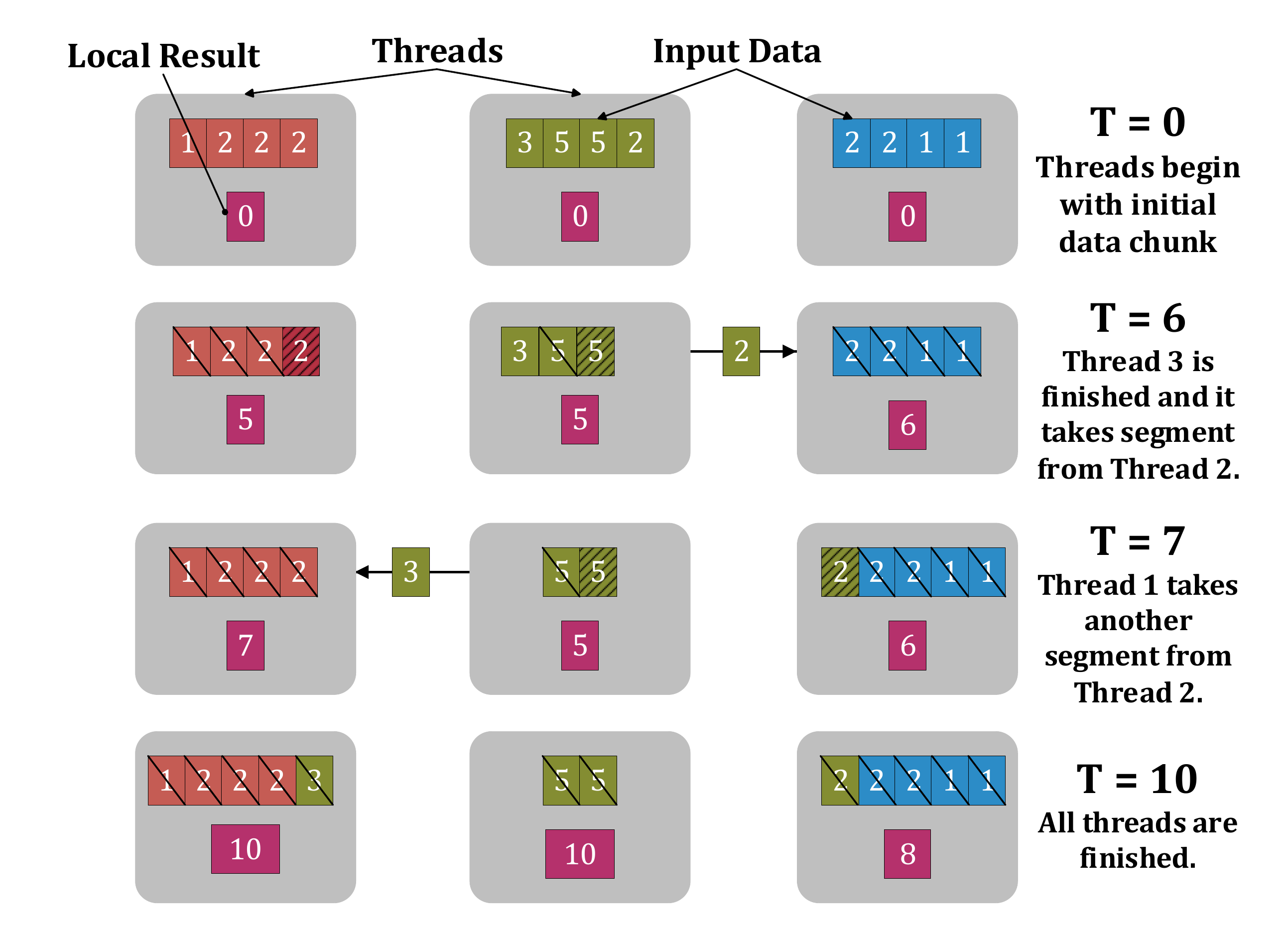}
	%\vspace{-5mm}
	\caption{An example of work-stealing in the computing reduction over three data segments. Data values indicate the computation time. Compared to the static data distribution, where the result of middle segment arrives after $t=15$ units of time, the computation is more balanced due to additional redistribution and the global phase is started earlier.}  \label{fig:load_balancing}
	%\vspace{-6mm}
\end{figure}

\subsection{Dynamic Hierarchical Prefix Sum}
\label{sec:dynamic_prefix_sum}

We now move away from one of the core assumptions that has been always made for a prefix scan operator - the computational cost is constant and easily predictable. Although this assumption is valid in many applications, it does not always hold as it is the case of image registration. If a static data distribution does not provide a balanced workload, the performance of the entire application is affected: not only it will take more time to process a single segment but the disparity will be later propagated due to a synchronous nature of the prefix scan.
% This is not strictly true.
%Obviously, for prefix scan algorithm it holds that rank $P'$ is not able to finish earlier than any of ranks $1, 2, \dots, P' -1$.
%
The ideally balanced data distribution can be estimated if each operator cost is known a priori which is not feasible for iterative computations with data-dependent stopping conditions.
A fine-grain work distribution is inefficient due to the increased depth of the global scan phase.
%Splitting data into more chunks of smaller size is less than ideal for prefix scan where a twofold increase in the number of data segments increases the depth by one.

Therefore, we focus on techniques that can react to ongoing changes in workload balance.
Well-known load balancing solutions do not apply directly to prefix scan due to the sequential nature
of the scan computation and its limited ability to redistribute work chunks.
%We have shown that our proposed algorithmic changes enable us to introduce flexibility to the static form of prefix scan and determine work distribution dynamically. The novelty of a more flexible scan structure is a crucial step for efficient prefix computations on irregular problems. To the best of our knowledge, no other work has presented work balancing in the context of the prefix scan.
%
We use the hierarchical prefix scan representation and introduce the shared--memory parallelism with
threads. We attempt to detect when a certain thread is processing its workload faster and let it steal
work from its neighbors to balance the computation effort. Thus, we improve the performance of global scan,
a main bottleneck in large scale computations, by (1) decreasing the work imbalance in first phase of computation
and (2) restricting the global phase only to parallel workers on the highest hierarchy levels, in our case MPI ranks.
The \textit{scan--then--map} approach imposes a strict evaluation order from left to right since otherwise
the new sequence would not contain a correct prefix scan. Fortunately, this requirement does not exist
in the \textit{reduce--then--scan} strategy where the first phase computes only a sum of the entire segment
$x_{l_{I}, r_{I}}$. Given the associativity of the operator, there shall be no change in result if
elements are processed from left to the right as in a prefix scan, from right to the left or from the middle of 
%the
data segment in both directions. This observations allows us to consider flexible segment boundaries.

An example of such problem is presented in Figure~\ref{fig:load_balancing}. The static data distribution leads to an unbalanced workload and effectively slows down the prefix scan to the slowest thread. By changing the order of evaluation to left-to-right in the lowest-numbered segment, right-to-left in the highest-numbered segment, and to middle-outward for other segments, we leave an option for each thread to acquire more work in case its neighbor is processing slower.
We note that there is no cost associated with changing segment boundaries since for the entire hierarchy it is only relevant that segment boundaries are aligned with each other.
%We note that there is no cost associated with changing segment boundaries - for the entire hierarchy of computation it does not matter how many elements are assigned to each segment. It only matters that segment boundaries are aligned with each other.
%
Any load balancing procedure will be restricted to exchanges between neighboring threads, due to the requirement that a sum must be computed across consecutive data elements. We focus on intra-node work-stealing due to diminishing returns of inter-node synchronization between logically adjacent threads. The main point of an efficient heuristic is to decide in which direction to accumulate data after starting on the middle element. Since the imbalance between neighbors cannot be predicted, the most sensible option is a greedy approach where threads always move in the direction of whichever adjacent thread is slower. Let $pl_{I}$ and $pr_{I}$ be the boundaries of processed elements for thread $I$.
%PPOPP
%and they are initialized after processing the middle element with corresponding position.
For each neighbor, we define the processing rate $t_{I \pm 1}$ as the ratio of computation time to the number of operator applications. Let $s_{I}$ correspond to the number of data elements left unprocessed between threads $I$ and $I+1$. The Algorithm~\ref{alg:load_balance} presents the heuristic. For simplicity, we omit the initial step where threads always move to the right.

%\vspace{-3mm}
\begin{algorithm}[htb]
	
	%\vspace{-3mm}
	\caption{Load balancing on threads $1, \dots, T$}
	\begin{algorithmic}[1]
		\While{$s_{I-1} > 0 \lor s_{I+1} > 0$}
		
		%\If{$t_{I-1} \le 0 \land t_{I+1} \le 0$} 
		%\State $d \gets RIGHT$ \Comment{First element finished}
		%\ElsIf{$t_{I-1} \le 0$} 
		%\State $d \gets LEFT$
		%\ElsIf{$t_{I+1} \le 0$} 
		%\State $d \gets RIGHT$
		%\Else
		\If{$s_{I-1} > 0 \land s_{I+1} > 0$}
			\If{$t_{I-1} > t_{I+1}$}
				\State $d \gets LEFT$
		\Else
			\State $d \gets RIGHT$
		\EndIf
		%\State \algorithmicif\ $t_{I-1} > t_{I+1}$ \algorithmicthen\ $d \gets LEFT$ \algorithmicelse\ $d \gets RIGHT$
		%\ElsIf{$s_{I-1} > 0$}
		%	\State $d \gets LEFT$
		%\Else
		%	\State $d \gets RIGHT$
		%\EndIf
		\Else
		\State \algorithmicif\ $s_{I-1} > 0$ \algorithmicthen\ $d \gets LEFT$ \algorithmicelse\ $d \gets RIGHT$
		%\If{$s_{I-1} > 0$}
		%	\State $d \gets LEFT$
		%\Else
		%	\State $d \gets RIGHT$
		%\EndIf
		\EndIf
		%\EndIf
		
		\If{$d == LEFT$}
		\State $pl_{I} \gets pl_{I} - 1$, \: $res_{I} \gets x_{pl_{I}} \odot res_{I}$
		\Else
        \State $pr_{I} \gets pr_{I} + 1$, \: $res_{I} \gets res_{I} \odot x_{pr_{I}}$
		\EndIf
		
		\EndWhile
	\end{algorithmic}
	\label{alg:load_balance}
	
	%\vspace{-5mm}
\end{algorithm}
%\vspace{-5mm}

\section{Evaluation}
\label{sec:evaluation}
%\review{We evaluate the effectiveness of our hierarchical, work-stealing prefix scan in the following manner: (1) a set of microbenchmarks to determine how our results on image registration translate to a generic prefix scan problem with unbalanced workload (\ref{sec:evaluation_benchmarks}), (2) the strong scaling of the two-step image registration (\ref{sec:strong_scaling}), (3) performance improvements of using a hierarchical prefix scan (\ref{sec:evaluation_hierarchical}), (4) impact of work--stealing prefix scan on energy consumption (\ref{sec:evaluation_energy}), and (5) weak scaling experiments to verify the ability of our dynamic prefix scan to process image sequences increasing in length (\ref{sec:evaluation_weak_scaling})}.
%We evaluate the effectiveness of our hierarchical, work-stealing prefix scan in the following manner: (1) a set of microbenchmarks to determine how our results on image registration translate to a generic prefix scan problem with unbalanced workload (\ref{sec:evaluation_benchmarks}), (2) the strong scaling of the two-step image registration (\ref{sec:strong_scaling}), (3) performance improvements of using a hierarchical prefix scan (\ref{sec:evaluation_hierarchical}), (4) impact of work--stealing prefix scan on energy consumption (\ref{sec:evaluation_energy}), and (5) weak scaling experiments to verify the ability of our dynamic prefix scan to process image sequences increasing in length (\ref{sec:evaluation_weak_scaling}).

For evaluation we use two supercomputing system: the Piz Daint supercomputer and a local cluster with Ivy Bridge CPUs, summarized in Table~\ref{tab:computingsystems}. We use 12 and 20 threads per rank on Piz Daint and IvyBridge, respectively, without hyper-threading and with each thread pined to a physical core. All prefix scan algorithms were implemented in C++ as a part of the quocmesh library~\cite{quocmesh}.
%No libraries other than an MPI implementation and OpenMP implementation provided with the compiler were used during experiments.
The work--stealing implementation splits the work across OpenMP threads and performs a local scan over partial results with the dissemination pattern since its implementation is simpler than a Ladner--Fischer scan and the difference in work performed is negligible when only a dozen or so threads participate in the scan. Images are available to all ranks through the high-performance filesystem and the communication is limited to 20 bytes of deformation data and indices. Algorithms use point-to-point communication with the exception of the Ladner-Fischer that uses \texttt{MPI\_Broadcast} in certain iterations. For each experiment, measurements were repeated five times and we show on plots mean value with 95\% confidence interval.

%\paragraph{Piz Daint} We used Cray XC50 nodes with 64 GB of system memory and a single, 12-core Intel Xeon E5-2690 CPU with base frequency 2.60GHz. Nodes are connected with each other by a Cray Aries interconnect with Dragonfly network topology. On this system, the source code was configured with CMake 3.5.2 and compiled with GCC 7.3.0. We use Cray MPICH in version 7.7.2.
%
%\paragraph{Ivy Bridge System} We use 31 nodes with 64 GB of system memory and two deca-core Intel Xeon E5-2680 v2 CPUs with base frequency 2.80 GHz. Nodes are connected with each other by a FDR Infiniband. The source code was configured with CMake 3.6.0 and compiled with GCC 8.2.0. We use IntelMPI 2018.3. 

%\subsubsection{Microscopic Data}

As image series, we consider data from an experiment where ultrahigh vacuum high--resolution TEM (UVH HRTEM) has been applied to capture the process of aluminum oxidation~\cite{doi:10.1021/acsami.7b17224}. The images have been acquired at a resolution of $\num{1920} \times \num{1856}$ and a rate of 400 frames per second.
%PPOPP
%, and each experiment has lasted for up to 4 minutes, producing up to $\num{96000}$ frames.
An example of a single frame is presented in the Figure~\ref{fig:electron_data_ref}.

\begin{table}[htb!]
	%\vspace{-3mm}
	\begin{adjustbox}{max width=\linewidth}
		\begin{tabular}{@{}l c c|c@{}}%\toprule
			&& Piz Daint & Ivy Bridge\\
			%\midrule
			
			\cmidrule{3-4}
			CPU & & Intel Xeon E5-2690 CPU 2.60GHz & Intel Xeon E5-2680 v2 2.80GHz \\
            Cores & &  12 with 12 hardware threads & 20 with 20 hardware threads \\
			Memory & & 64 GB & 64 GB \\
			Interconnect & & Cray Aries, Dragonfly & FDR Infiniband \\
			Build & & CMake 3.5.2, GCC 7.3.0 & CMake 3.6.0, GCC 8.2.0 \\
			MPI & & Cray MPICH 7.7.2 & IntelMPI 2018.3
		\end{tabular}
		%\hfill
		%		\begin{tabular}{@{}rc cccc c cccc@{}}%\toprule
		%			
		%			\multicolumn{11}{c}{Work--stealing} \\
		%			&&\multicolumn{4}{c}{Dissemination} && \multicolumn{4}{c}{Ladner--Fischer} \\
		%			\cmidrule{3-6} \cmidrule{8-11}
		%			
		%			Cores &\phantom{a}& \multicolumn{2}{c}{Work} & \multicolumn{2}{c}{Energy [\si{\mega\joule}]} & \phantom{a} & \multicolumn{2}{c}{Work} & \multicolumn{2}{c}{Energy [\si{\mega\joule}]}  \\
		%			%\midrule
		%			
		%			\cmidrule{1-1} \cmidrule{3-6} \cmidrule{8-11}
		%			%$64$ & 0.0790 && 0.3670 & 0.7187 & 3.1815 & 1&& -1.0032 & -1.7104 & -21.7969&4\\
		%			%$128$ & -0.8651&& -9.0714& 297.0923& 46.2143 & 2&& 4.3590& 34.5809& 76.9167&5\\
		%			%$c$ & 124.27561&& 128.2265& -630.5455& -381.0930& 3&& -121.0518& -137.1210& -220.2500&6\\
		%			\csvreader[head to column names, late after line=\\,]{energydataOMP.csv}{}% use head of csv as column names
		%			{\ranks && \workKS & \workincreaseKS & \energyKS & \energyincreaseKS && \workLF & \workincreaseLF & \energyLF & \energyincreaseLF}
		%			%\bottomrule	
		%		\end{tabular}
	\end{adjustbox}
	\caption{Evaluation systems: Piz Daint with Cray XC50 nodes and Ivy Bridge cluster with two deca-core CPUs.}
	\label{tab:computingsystems}
	
	%\vspace{-0.8cm}
\end{table}

%\begin{figure*}[htb]
%	\centering
%	\subfloat[Prefix operator with constant runtime.]{%
%		\includegraphics[width=0.49\textwidth]{images/mock_runs_static.pdf}\label{fig:mock_runs_static}}
%	\hfill
%	\subfloat[Prefix operator with runtime given by exponential distribution with rate $\lambda$.]{%
%		\includegraphics[width=0.49\textwidth]{images/mock_runs_dynamic.pdf} \label{fig:mock_runs_dynamic}}
%	\label{fig:mock_runs}
%	\caption{Comparison of prefix scan algorithms on operators with a static and dynamic running time. Allocation of one MPI rank per node.}
%\end{figure*}
%
%\begin{figure}[htb]
%	\centering
%	\includegraphics[width=\linewidth]{images/microbenchmarks_dynamic.pdf}
%	\caption{\hll{Distributed and work--stealing prefix scan on operator with a dynamic running time.}}
%	\label{fig:microbenchmarks_dynamic}
%\end{figure}

\subsection{Microbenchmarks}
\label{sec:evaluation_benchmarks}

We first evaluate the prefix scan algorithms with a set of microbenchmarks.
We use an artificial operator with (1) a static execution time, where each operator application takes the same amount of time and (2) a dynamic configuration where the execution time is a random variable and for each time $t$, we use an exponential distribution with rate $\lambda = \frac{1}{t}$ to obtain a similar average running time. We use \texttt{std::mt19937}, a 32-bit Mersenne Twister PRNG from the C++ standard library, with a constant seed 1410 to ensure reproducible results. We scale it up on Piz Daint with varying number of data elements per CPU core. Whenever we compare static and work--stealing implementations, both solutions use random number generators in the same deterministic fashion to ensure that the comparison is scientifically valid. We evaluate (1) the scalability of prefix scan algorithms and (2) effectiveness of work-stealing on large and generic problems with an unbalanced workload.
%PPOPP
%We use an artificial operator with a predefined execution running time and scale it up on Piz Daint, simulating a scenario with varying number of data elements per CPU core. We design two timing configurations: (1) a static execution time, where each operator application takes the same amount of time and (2) a dynamic configuration where the execution time is a random variable and for each time $t$, we use an exponential distribution with rate $\lambda = \frac{1}{t}$ to obtain a similar average running time. We use \texttt{std::mt19937}, a 32-bit Mersenne Twister PRNG from the C++ standard library. To ensure reproducible results, the generator is seeded with a constant 1410. Whenever we compare static and work--stealing implementations, both solutions use random numbers in a same fashion to ensure that the comparison is scientifically valid. We evaluate (1) the scalability of prefix scan algorithms and (2) effectiveness of work-stealing on large scale and generic problems with an unbalanced workload.
%\vspace{-2mm}

\begin{figure}[htb]
	%\vspace{-3mm}
	\centering
	\subfloat[Prefix operator with constant runtime.]{%
		\includegraphics[width=\linewidth]{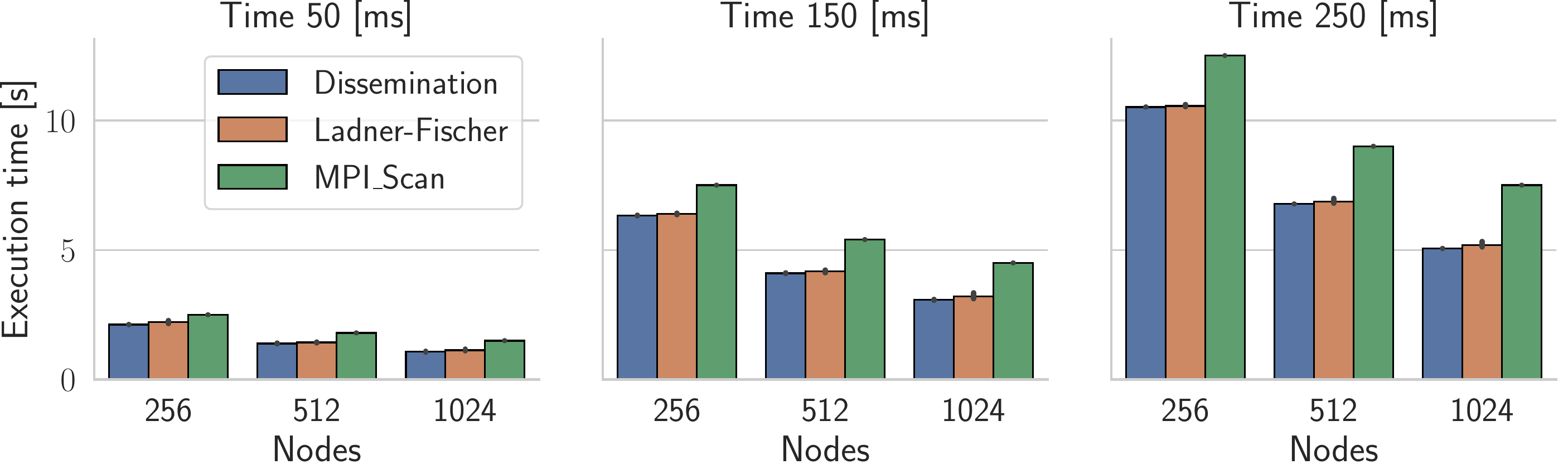}\label{fig:mock_runs_static}}	%\hfill
	\newline
	\subfloat[Prefix operator with exponentially distributed time with rate $\lambda$.]{%
		\includegraphics[width=\linewidth]{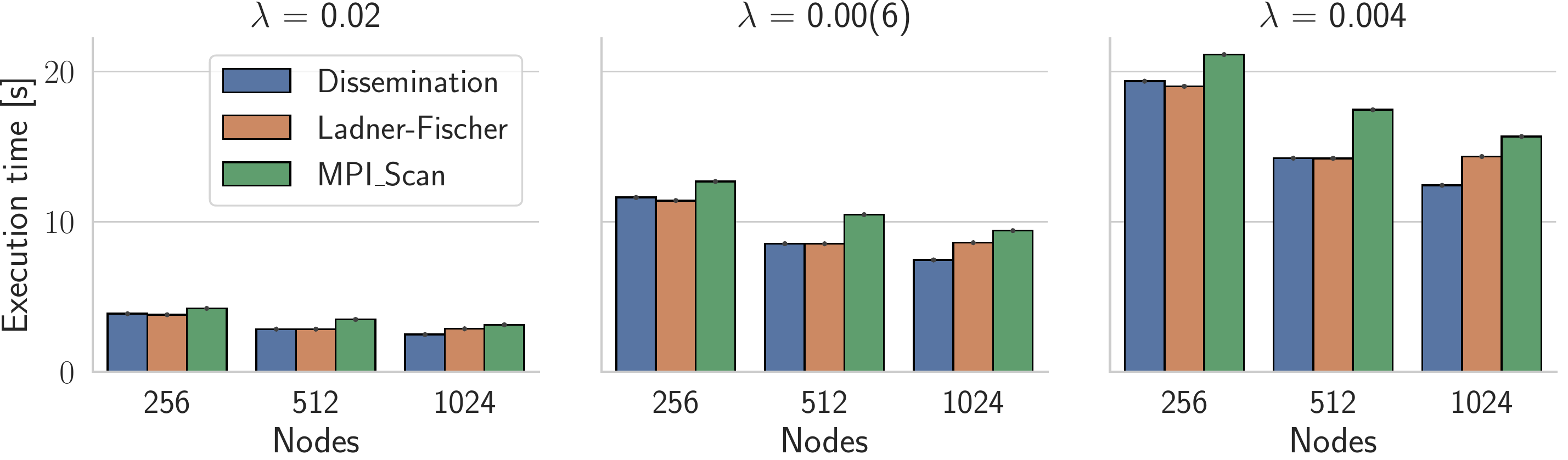} \label{fig:mock_runs_dynamic}}
	\newline
	\subfloat[Work--stealing prefix scan with a dynamic execution time.]{%
			\includegraphics[width=\linewidth]{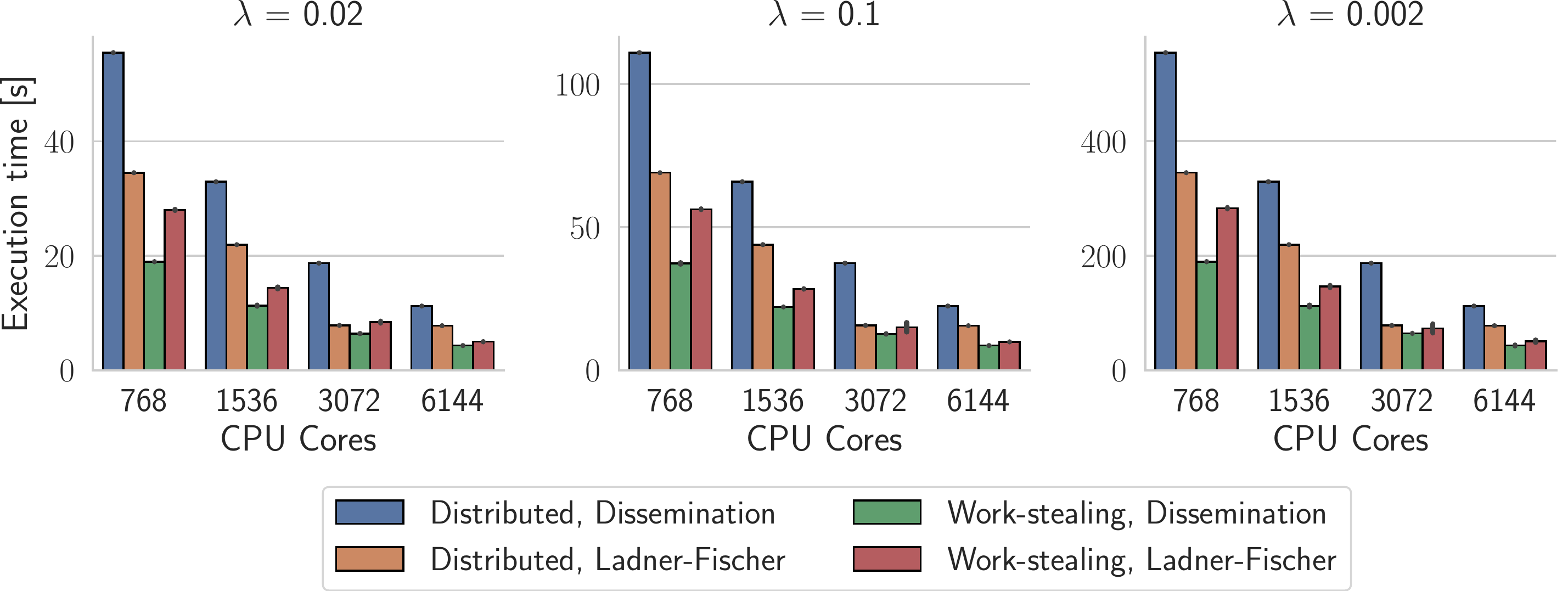} \label{fig:microbenchmarks_dynamic}}
	\label{fig:mock_runs}
	%\vspace{-3mm}
	\caption{Prefix scan algorithms on mock operators with a static and dynamic running time.}
	
	%\vspace{-6mm}
\end{figure}

\begin{figure}[htb]
	
	%\vspace{-5mm}
		\includegraphics[width=\dimexpr0.49\textwidth]{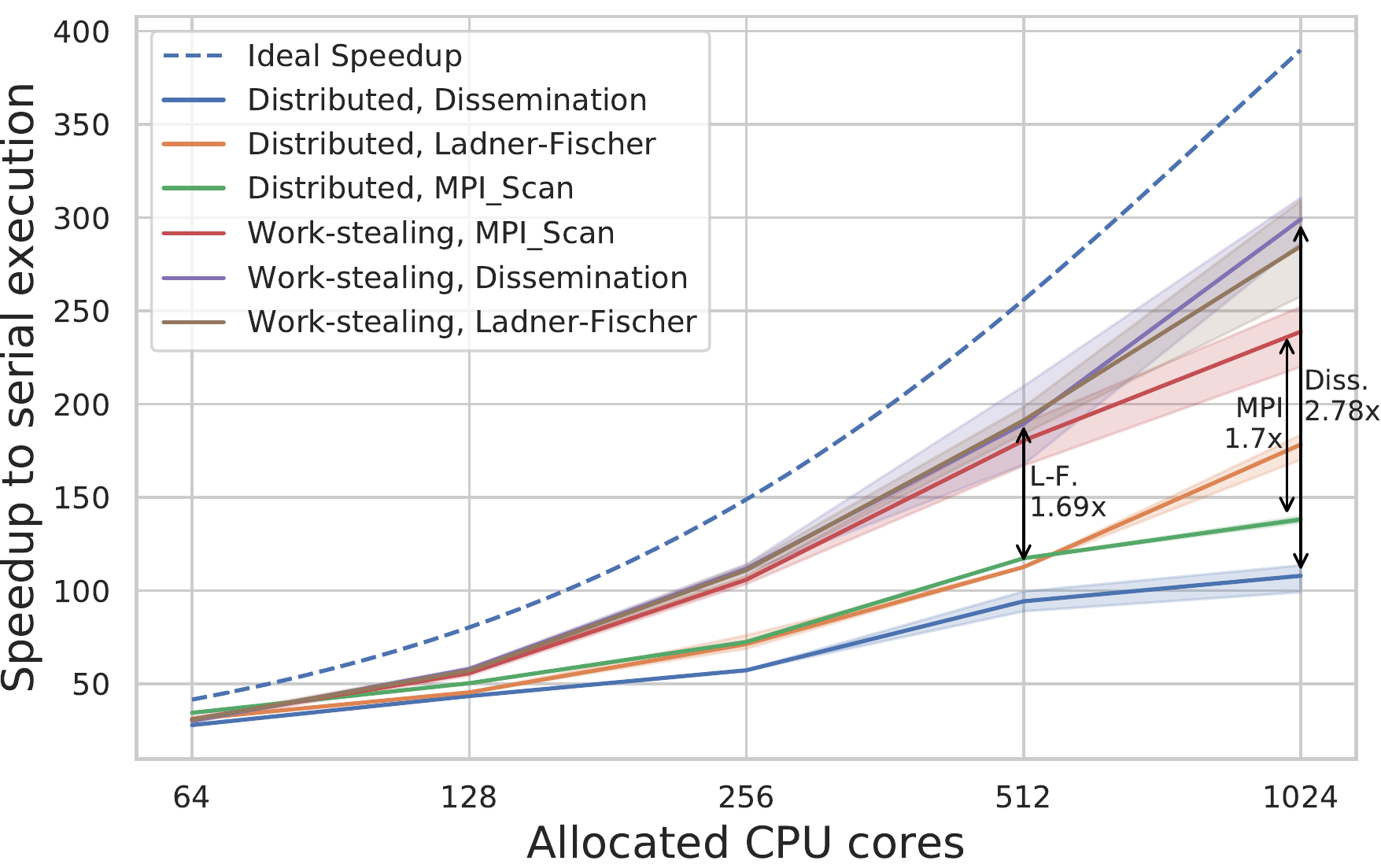} 
	\centering 
	%	\caption{[Insert an overview/introductory figure. Suggestion: example
	%  of microscopy images, present as series and visualize scheduling of
	%  prefix sum on cluster.]\htor{needs main result figure - speedup or so}}
	
	%\caption{The strong scaling of prefix scan algorithms on the \textit{scan} image registration. Experimental results are indicated with solid lines whereas theoretical bound~\eqref{eq:scan_speedup} is presented with dashed lines. Experiments were conducted on the Piz Daint system for $\num{4096}$ images.} 
	\caption{The strong scaling of \textit{full} image registration. Experimental results are indicated with solid lines whereas theoretical bound~\eqref{eq:full_speedup} discussed in Section~\ref{sec:strong_scaling} is dashed. Experiments conducted on Piz Daint for $\num{4096}$ images.}
	\label{fig:strong_full}
	%\resizebox*{0.9\width}{0.8\totalheight}{
	%}
	%\vspace{-6mm}
\end{figure}

\subsubsection{Inter-Node Scan}

We test sensitivity of scan algorithms to unbalanced workloads and network congestion.
We estimate the performance loss on different implementations of global scan by using a static,
hierarchical prefix scan with one MPI rank and 12 threads per node.
Figures~\ref{fig:mock_runs_static},~\ref{fig:mock_runs_dynamic} present results for static and dynamic execution with 98304 data elements. Results show that
scan algorithms perform differently on an ideally constant workload if the
computation time plays a more important role than communication. Not surprisingly,
the \texttt{MPI\_Scan} performs worse than other prefix
scan algorithms since it might be optimized for communication latency.
Adding a controlled imbalance causes a performance drop and all prefix scan algorithms take on
average twice more time. We can expect such slowdown in the image
registration and other problems where load balance is an issue.

%\vspace{-2mm}

%PPOPP
%We test sensitivity of scan algorithms to unbalanced workloads and network congestion. We assume a static hierarchical prefix scan with one MPI rank per node and 12 threads per node. This way we can estimate the performance loss caused by load imbalance on different implementations of global scan.
%Figure~\ref{fig:mock_runs_static} presents results for static and dynamic execution with 98304 data elements. Results show that
%scan algorithms perform differently on an ideally constant workload if the
%computation time is increased and the communication plays a less important
%role. Not surprisingly, the \texttt{MPI\_Scan} performs worse than other prefix
%scan algorithms since it might be optimized for communication latency.
%Figure~\ref{fig:mock_runs_dynamic} shows that adding a controlled imbalance
%causes a performance drop with all prefix scan algorithms taking on
%average twice more time and \texttt{MPI\_Scan} providing
%the slowest execution time. We can expect such slowdown in the image
%registration and other prefix scan problems where load balance is an issue.

\subsubsection{Work-stealing Scan}

We evaluate the impact of our work--stealing on a generic prefix scan problem with an unbalanced workload.
Figure~\ref{fig:microbenchmarks_dynamic} presents results for dynamic execution with 98304 data elements.
Results show that our work--stealing provides substantial improvements when applied with the Ladner--Fischer scan
whereas the performance of scan with the dissemination can be improved up to three times. The result is
explained by chains of dependencies in task graphs of both algorithms.
Dissemination and Ladner--Fischer scans represent distinct sequences of parallel computations,
and the critical path is different in both algorithms as well, resulting in different impact of workload
imbalance.
The performance seems to be consistent across operators with varying execution time,
%and the distribution of imbalance is a primary factor affecting the performance. Furthermore,
and we see that work-stealing prefix scan improves the performance on a larger number of cores even if the distributed version stops to scale, as it
is in the case of Ladner--Fischer from 3072 to 6144 cores.

% PPOPP - remove table
\begin{table*}\centering

\begin{adjustbox}{max width=\textwidth}
	\begin{tabular}{@{}rc SSSSSSSSS c SSSSSSSSS @{}}%\toprule
		\multicolumn{21}{c}{\emph{Scan} Registration} \\
		& \phantom{ab}& \multicolumn{9}{c}{Distributed} &
		\phantom{ab} & \multicolumn{9}{c}{Work--stealing}\\
		%\cmidrule{2-2} \cmidrule{4-7} \cmidrule{9-13}
		& & \multicolumn{3}{c}{Dissemination} & \multicolumn{3}{c}{Ladner--Fischer} & \multicolumn{3}{c}{\code{MPI\_Scan}} 
		&&\multicolumn{3}{c}{Dissemination} & \multicolumn{3}{c}{Ladner--Fischer} & \multicolumn{3}{c}{\code{MPI\_Scan}} \\
		\cmidrule{1-1} \cmidrule{3-11} \cmidrule{13-21}
		Cores && {Time} &{$\mathcal{S}$}& {$\mathcal{E}$} & {Time} & {$\mathcal{S}$} & {$\mathcal{E}$} & {Time} & {$\mathcal{S}$} & {$\mathcal{E}$} && {T} & {$\mathcal{S}$} & {$\mathcal{E}$} & {Time} & {$\mathcal{S}$} & {$\mathcal{E}$} & {Time} & {$\mathcal{S}$} & {$\mathcal{E}$} \\
		
		\cmidrule{1-1} \cmidrule{3-11} \cmidrule{13-21}
		\begin{filecontents*}{tabledataScan.csv}
ranks,distrKST,distrKSS,distrLFT,distrLFS,distrMPIT,distrMPIS,staticKST,staticKSS,staticLFT,staticLFS,staticMPIT,staticMPIS,ompKST,ompKSS,ompLFT,ompLFS,ompMPIT,ompMPIS,distrKSE,distrLFE,distrMPIE,staticKSE,staticLFE,staticMPIE,ompKSE,ompLFE,ompMPIE
64,979.2603999999999,18.8123267995588,765.6058,24.062208863447307,779.8181999999999,23.623668525134025,683.4345999999999,26.955273652616754,682.799,26.980365622484317,729.2090000000001,25.26321900397097,780.8933999999999,23.59114146267169,805.3696,22.87417685826069,800.1314,23.023926653380517,0.29394260624310625,0.37597201349136417,0.36911982070521915,0.4211761508221368,0.42156821285131746,0.3947377969370464,0.36861158535424515,0.3574090134103233,0.3597488539590706
128,677.376,27.196367551650294,403.999,45.599535312381136,552.2234,33.35998921209545,403.2558,45.683575206275194,402.62019999999995,45.75569399316445,437.8278,42.07628356780147,455.4348,40.44962454925857,473.4264,38.91241947357956,466.08680000000004,39.52518429328328,0.21247162149726792,0.3562463696279776,0.2606249157194957,0.35690293129902495,0.35746635932159726,0.32872096537344897,0.3160126917910826,0.3040032771373403,0.30879050229127564
256,550.9046000000001,33.43984905311494,338.418,54.436131253853716,398.262,46.25640072782909,274.5114,67.10893123807124,329.97639999999996,55.828740075552886,300.5968,61.285305321502655,286.51939999999996,64.29640250072654,260.9522,70.59594311397515,288.2458,63.911309953750134,0.13062441036373024,0.21264113771036608,0.18068906534308238,0.26214426264871576,0.21808101592012846,0.23939572391211975,0.25115782226846306,0.27576540278896544,0.24965355450683646
512,340.755,54.062791937511314,283.5598,64.96748363719634,298.3236,61.75229404132515,202.38899999999998,91.02355694561794,244.7312,75.27510455008053,243.26659999999998,75.7283024741854,171.6014,107.3544077534721,186.877,98.57910104863984,193.10639999999998,95.39904770979454,0.10559139050295178,0.1268896164788991,0.12060994929946318,0.17778038465941004,0.14702168857437603,0.14790684076989336,0.2096765776435002,0.1925373067356247,0.18632626505819247
1024,274.454,67.12296656877534,166.1062,110.90595454394037,308.2688,59.76007518979108,162.526,113.34904364019705,215.74260000000004,85.38956453971846,175.54850000000002,104.94060995489376,143.5754,128.3100494002919,153.0622,120.35738847779967,167.935,109.69819672293845,0.06554977203981967,0.10830659623431677,0.05835944842753035,0.11069242542987993,0.08338824662081881,0.10248106440907594,0.12530278261747255,0.11753651218535124,0.10712714523724458


\end{filecontents*}
\csvreader[head to column names, late after line=\\,]{tabledataScan.csv}{}% use head of csv as column names
		{\ranks && \distrKST & \distrKSS & \distrKSE & \distrLFT & \distrLFS & \distrLFE & \distrMPIT & \distrMPIS & \distrMPIE && \ompKST & \ompKSS  & \ompKSE & \ompLFT & \ompLFS & \ompLFE & \ompMPIT & \ompMPIS  & \ompMPIE}% specify your coloumns here
		\\
	\end{tabular}
\end{adjustbox}

%\vspace{-3mm}

\begin{adjustbox}{max width=\textwidth}
\begin{tabular}{@{}rc SSS SSS SSS  c SSSSSSSSS @{}}%\toprule
	\multicolumn{21}{c}{\emph{Full} Registration} \\
	& \phantom{ab}& \multicolumn{9}{c}{Distributed} &
	\phantom{ab} & \multicolumn{9}{c}{Work--stealing}\\
	%\cmidrule{2-2} \cmidrule{4-7} \cmidrule{9-13}
	& & \multicolumn{3}{c}{Dissemination} & \multicolumn{3}{c}{Ladner--Fischer} & \multicolumn{3}{c}{\code{MPI\_Scan}}
	&&\multicolumn{3}{c}{Dissemination} & \multicolumn{3}{c}{Ladner--Fischer} & \multicolumn{3}{c}{\code{MPI\_Scan}} \\
	\cmidrule{3-11}  \cmidrule{13-21}
	Cores && {Time} &{$\mathcal{S}$}& {$\mathcal{E}$} & {Time} & {$\mathcal{S}$}& {$\mathcal{E}$} & {Time} & {$\mathcal{S}$}& {$\mathcal{E}$} && {Time} & {$\mathcal{S}$} & {$\mathcal{E}$} & {Time} & {$\mathcal{S}$} & {$\mathcal{E}$} & {Time} & {$\mathcal{S}$} & {$\mathcal{E}$} \\
	
	\cmidrule{1-1}	\cmidrule{3-11}  \cmidrule{13-21}
	\begin{filecontents*}{tabledataFull.csv}
ranks,distrKST,distrKSS,distrLFT,distrLFS,distrMPIT,distrMPIS,staticKST,staticKSS,staticLFT,staticLFS,staticMPIT,staticMPIS,ompKST,ompKSS,ompLFT,ompLFS,ompMPIT,ompMPIS,distrKSE,distrLFE,distrMPIE,staticKSE,staticLFE,staticMPIE,ompKSE,ompLFE,ompMPIE
64,1345.5120000000002,27.920746897835173,1203.956,31.203548966905775,1088.5,34.51327514928801,1048.81,35.8193571762283,1048.296,35.8369201065348,1108.24,33.898523785461634,1247.1,30.124047790874837,1222.255,30.736384796953175,1215.455,30.908342966214303,0.4362616702786746,0.48755545260790273,0.5392699242076252,0.5596774558785672,0.5599518766646062,0.529664434147838,0.4706882467324193,0.48025601245239335,0.4829428588470985
128,864.0106,43.48060081670295,826.5368000000001,45.45193874005368,745.6882,50.379904093963134,593.198,63.330793428163965,593.6812000000001,63.279248189095426,630.267,59.60600824729837,648.4670000000001,57.933094513676096,657.33375,57.15163720104133,674.2436666666666,55.71828384496012,0.3396921938804918,0.35509327140666935,0.393593000734087,0.494771823657531,0.494369126477308,0.4656719394320185,0.4526023008880945,0.4464971656331354,0.43529909253875093
256,655.4130000000001,57.31912549796845,529.1253333333333,70.99962453760169,518.3567999999999,72.47459664848616,377.79139999999995,99.44032606353667,433.70640000000003,86.62011904827783,404.4848,92.87790295210105,336.0786,111.7824818360943,338.5446,110.96824465668631,355.2097,105.76203296250075,0.22390283397643926,0.2773422833500066,0.28310389315814904,0.38843877368569013,0.33835984003233527,0.36280430840664474,0.4366503196722434,0.4334697056901809,0.41313294125976857
512,400.26539999999994,93.8569758964927,333.4306,112.67022282897851,320.1002,117.36231342560863,269.6962,139.29636383456648,311.8274,120.47594278116676,310.6456,120.93427365460835,202.59439999999998,185.43306231564154,196.8096,190.88347316391074,211.13809999999998,177.92951627394586,0.1833144060478373,0.22005902896284865,0.22922326840939186,0.27206321061438765,0.23530457574446634,0.23619975323165693,0.3621739498352374,0.37281928352326316,0.3475185864725505
1024,350.4386,107.20194636093171,211.0324,178.0186360009174,271.9022,138.16622300224125,204.7548,183.47652899956438,258.18379999999996,145.50758025871497,218.09125,172.25679618049787,125.8398,298.5359162999306,133.54940000000002,281.30190027061144,160.08689999999999,234.67066949263187,0.10468940074309738,0.1738463242196459,0.13492795215062622,0.1791762978511371,0.14209724634640133,0.16821952752001745,0.29153898076165097,0.274708886983019,0.2291705756763983

\end{filecontents*}
\csvreader[head to column names, late after line=\\,]{tabledataFull.csv}{}% use head of csv as column names
	{\ranks && \distrKST & \distrKSS & \distrKSE & \distrLFT & \distrLFS & \distrLFE & \distrMPIT & \distrMPIS & \distrMPIE && \ompKST & \ompKSS & \ompKSE & \ompLFT & \ompLFS& \ompLFE& \ompMPIT & \ompMPIS & \ompMPIE}% specify your coloumns here
	%\bottomrule
\end{tabular}
\end{adjustbox}

	\caption{Execution times, parallel speedups $\mathcal{S}$ and efficiency $\mathcal{E}$ for (a) the standard, MPI-only distributed prefix scan, (b) ours hierarchical prefix scan with MPI ranks and work-stealing on OpenMP threads. Speedups are computed relative to the serial \emph{scan} and \emph{full} registration lasting 18422.17 and 37567.7, respectively.}
	\label{tab:scaling_data}
	
	%\vspace{-8mm}
\end{table*}

\subsection{Strong Scaling}
\label{sec:strong_scaling}

We evaluate the strong scaling of the image registration on the Piz Daint system. As a baseline we
choose the serial execution of a prefix scan that requires $N - 1$ operator applications for $N$
deformations $\phi_{i, j}$. On a single core, this step takes on average 18422 seconds. Combined
with the depth of a distributed prefix scan (Section~\ref{sec:distributed_prefix_scan2}, Eq.~(\ref{eq:reduce_then_scan})),
we obtain the upper performance bound for \textit{scan registration}
\begin{align}
\dfrac{ N - 1 }{D_{DS}(N, P)} &= \dfrac{N - 1}{2 \cdot \frac{N}{P} - 1 + C_{1} \log_2{P}}
\label{eq:scan_speedup}
\end{align}
Figure~\ref{fig:strong_scan2} and Table~\ref{tab:scaling_data}
present parallel speedups and efficiency. In addition, in the Figure we compare with the upper performance bound although this can be achieved only on perfectly balanced workloads.
For the \textit{scan} computation, we observe that our work-stealing prefix scan up to $1.98$x, $1.83$x and $1.51$x times over the dissemination, \texttt{MPI\_Scan} and the Ladner--Fischer scan, respectively. 
%PPOPP
%For the \textit{scan} computation, we observe that our work-stealing prefix scan improves the performance up to 1.98x times over the dissemination prefix scan, up to 1.83x times over the Cray MPICH implementation of \texttt{MPI\_Scan} and up to 1.51x times over the Ladner--Fischer scan.
%
% PPOPP
%Applying our load balancing helps to achieve scaling closer to the performance bound for the prefix scan problem. It prevents the \texttt{MPI\_Scan} algorithm from stopping to scale that is observed for the run on $1024$ cores with only 4 images per core.
Applying our load balancing brings the performance closer to the upper bound and prevents the \texttt{MPI\_Scan} algorithm from stopping to scale on $1024$ cores.
%Furthermore, the work--efficient Ladner-Fischer scan provides the best performance on a large scale and continues to improve 
The work--efficient Ladner--Fischer exhibits a substantial performance improvement on $1024$ cores. This is explained by the fact that scan algorithms are differently affected by various workload imbalances and the scan hits the sweet-spot for this configuration.
We don't observe any significant performance improvements over 512 cores on dissemination and \texttt{MPI\_Scan} whereas the work--stealing version provides further improvements. In turn, the dynamic prefix scan allows further scaling of long image series registration.

Minor slowdowns observed for some algorithms on $64$ or $128$ ranks can be an effect of measurement
noise that is noticeably larger for the dynamic execution. At the same time, we observe significant
performance improvements from 512 cores onwards. Both results align with the
analysis presented in Section~\ref{sec:load_imbalance} where it has been shown that negative effects
of an imbalanced operator have the highest impact when the local segment size is small.
When the number of allocated cores is relatively small, and the data segment is in turn large, applying
work-stealing on lower levels of hierarchical prefix scan cannot prevent all effects of unbalanced
workload on the highest level, which is the entire node in our setup. Furthermore, in such setup, the
computation is not dominated by the global scan which is especially sensitive to imbalanced workloads.

In addition, we study the performance of a \textit{full registration} that includes the initial step of generating input deformations for the prefix scan (Section~\ref{sec:image_registration}, function \textbf{A}).
A full serial registration requires on average 37567 seconds of computation. The upper
performance bound Eq.~(\ref{eq:scan_speedup}) is changed by adding $N$ initial registration steps that are massively parallel, adding depth $\frac{N}{P}$ on $P$ processes.
%\vspace{-2mm}
\begin{align}
  \dfrac{ \mathbf{N} + N - 1 }{\mathbf{\frac{N}{P}} + D_{DS}(N, P)} &= \dfrac{2N - 1}{3 \cdot \frac{N}{P} - 1 + C_{1} \log_2{P}}
\label{eq:full_speedup}
\end{align}
Figure~\ref{fig:strong_full} and Table~\ref{tab:scaling_data}
present results for the \textit{full} registration. We observe that our work-stealing prefix scan improves the performance up to $2.78$x, $1.7$x and $1.69$x times over the dissemination, \texttt{MPI\_Scan} and the Ladner--Fischer scan, respectively. 
Similar to the \emph{scan} registration, the work--stealing scan prevents stopping to scale over 512 cores when using dissemination and \texttt{MPI\_Scan}. We observe a substantial improvement on dissemination prefix scan. Although more work is performed in first phase, the time spent in global scan phase decreases due to lower waiting times. Such result is not surprising since different scan algorithms are affected differently by load imbalance. Furthermore, our work-stealing performs better when more work is performed by each thread.
%\vspace{-3mm}
%We use scaling bounds in~\eqref{eq:scan_speedup} and~\eqref{eq:full_speedup} to determine the parallel efficiency. 

%\begin{figure*}[htb]
%	\centering
%	
%	\subfloat[Comparison of prefix sum algorithms on a static and dynamic sum operator. Allocation of one MPI rank per node.]{
%		\includegraphics[width=\dimexpr0.49\textwidth]{images/weak_scaling.pdf} \label{fig:weak_scan}}\hfill
%	%\hfill
%	\subfloat[Comparison of distributed prefix sum algorithm with dynamic prefix sum.]{
%		\includegraphics[width=\dimexpr0.49\textwidth]{images/weak_scaling.pdf} \label{fig:weak_full}}\hfill
%	
%	\caption{The weak scaling of prefix scan algorithms on the image registration. Experiments were conducted on the Ivy Bridge system for 4096 images. TODO: I could draw it without confidence-intervals but as a stacked barplot with different phases (local-global-local)}  \label{fig:weak_scaling}
%\end{figure*}

% PPOPP 
\begin{table}
	%\centering
\begin{adjustbox}{max width=\linewidth}
	\begin{tabular}{@{}rc SSS  SSS  SSS @{}}%\toprule
		\multicolumn{11}{c}{Hierarchical \emph{Scan} Registration} \\
		
		& & \multicolumn{3}{c}{Dissemination} & \multicolumn{3}{c}{Ladner--Fischer} & \multicolumn{3}{c}{\code{MPI\_Scan}} \\
		\cmidrule{3-11} 
		Cores && {Time} &{$\mathcal{S}$} & {$\mathcal{S'}$} & {Time} &{$\mathcal{S}$} & {$\mathcal{S'}$}  & {Time} &{$\mathcal{S}$} & {$\mathcal{S'}$}  \\
		
		\cmidrule{1-1}	\cmidrule{3-11}
		%$64$ & 0.0790 && 0.3670 & 0.7187 & 3.1815 & 1&& -1.0032 & -1.7104 & -21.7969&4\\
		%$128$ & -0.8651&& -9.0714& 297.0923& 46.2143 & 2&& 4.3590& 34.5809& 76.9167&5\\
		%$c$ & 124.27561&& 128.2265& -630.5455& -381.0930& 3&& -121.0518& -137.1210& -220.2500&6\\
		\begin{filecontents*}{tableDataStatic.csv}
ranks,staticKST,staticKSS,staticKSSD,staticKSSW,staticLFT,staticLFS,staticLFSD,staticLFSW,staticMPIT,staticMPIS,staticMPISD,staticMPISW
64,683.4345999999999,26.955273652616754,1.4328516583737492,1.142601501299466,682.799,26.980365622484317,1.121275514463261,1.1795119793672808,729.2090000000001,25.26321900397097,1.0694028735245997,1.0972593591137794
128,403.2558,45.683575206275194,1.6797675321718868,1.1293942951347506,402.62019999999995,45.75569399316445,1.0034245673714337,1.1758635061032707,437.8278,42.07628356780147,1.261279891318002,1.0645436402165418
256,274.5114,67.10893123807124,2.0068550887139844,1.0437431742361154,329.97639999999996,55.828740075552886,1.0255824355923637,0.7908207980934395,300.5968,61.285305321502655,1.324904323665455,0.9589117382487106
512,202.38899999999998,91.02355694561794,1.6836636378459304,0.8478790843375876,244.7312,75.27510455008053,1.1586581522911668,0.7636010447380637,243.26659999999998,75.7283024741854,1.2263237123386441,0.7938056436847475
1024,162.526,113.34904364019705,1.6886775039070672,0.8833995791442599,215.74260000000004,85.38956453971846,0.7699276823399736,0.7094667441664277,175.54850000000002,104.94060995489376,1.7560320936949045,0.9566302189993078
\end{filecontents*}
\csvreader[head to column names, late after line=\\,]{tableDataStatic.csv}{}% use head of csv as column names
		{\ranks && \staticKST & \staticKSS & \staticKSSD  & \staticLFT & \staticLFS & \staticLFSD &  \staticMPIT & \staticMPIS & \staticMPISD }% specify your coloumns here
		%\bottomrule
	\end{tabular}
\end{adjustbox}
\caption{ Execution times, parallel speedups $\mathcal{S'}$ and $\mathcal{S'}$ with respect to serial and distributed execution for hierarchical prefix scan without work--stealing.}
\label{tab:static_data}

%\vspace{-8mm}
\end{table}
\begin{table}\centering
	\begin{adjustbox}{max width=\linewidth}
		\begin{tabular}{@{}rc cccc c cccc@{}}%\toprule
			\multicolumn{11}{c}{Distributed} \\
			&&\multicolumn{4}{c}{Dissemination} && \multicolumn{4}{c}{Ladner--Fischer} \\
			\cmidrule{3-6} \cmidrule{8-11}
			
			Cores &\phantom{a}& \multicolumn{2}{c}{Work} & \multicolumn{2}{c}{Energy [\si{\mega\joule}]} & \phantom{a} & \multicolumn{2}{c}{Work} & \multicolumn{2}{c}{Energy [\si{\mega\joule}]}  \\
			%\midrule
			
			\cmidrule{1-1} \cmidrule{3-6} \cmidrule{8-11}
			%$64$ & 0.0790 && 0.3670 & 0.7187 & 3.1815 & 1&& -1.0032 & -1.7104 & -21.7969&4\\
			%$128$ & -0.8651&& -9.0714& 297.0923& 46.2143 & 2&& 4.3590& 34.5809& 76.9167&5\\
			%$c$ & 124.27561&& 128.2265& -630.5455& -381.0930& 3&& -121.0518& -137.1210& -220.2500&6\\
			\begin{filecontents*}{energydataMPI.csv}
,ranks,energyKS,energyincreaseKS,workKS,workincreaseKS,energyLF,energyincreaseLF,workLF,workincreaseLF
0,64,1.34$\pm$0.12,1.89x,12545,1.53x,1.18$\pm$0.14,1.66x,12392,1.51x
1,128,1.76$\pm$0.1,2.49x,12929,1.58x,1.16$\pm$0.13,1.64x,12529,1.53x
2,256,2.34$\pm$0.19,3.3x,13825,1.69x,1.74$\pm$0.13,2.45x,12824,1.57x
3,512,3.3$\pm$0.36,4.65x,15873,1.94x,2.85$\pm$0.13,4.01x,13448,1.64x
4,1024,4.92$\pm$0.21,6.94x,20481,2.5x,3.5$\pm$0.22,4.94x,14751,1.8x
\end{filecontents*}
\csvreader[head to column names, late after line=\\,]{energydataMPI.csv}{}% use head of csv as column names
			{\ranks && \workKS & \workincreaseKS & \energyKS & \energyincreaseKS && \workLF & \workincreaseLF & \energyLF & \energyincreaseLF}  % specify your coloumns here
			
			\multicolumn{11}{c}{Work--stealing} \\
			&&\multicolumn{4}{c}{Dissemination} && \multicolumn{4}{c}{Ladner--Fischer} \\
			\cmidrule{3-6} \cmidrule{8-11}
			
			Cores &\phantom{a}& \multicolumn{2}{c}{Work} & \multicolumn{2}{c}{Energy [\si{\mega\joule}]} & \phantom{a} & \multicolumn{2}{c}{Work} & \multicolumn{2}{c}{Energy [\si{\mega\joule}]}  \\
			%\midrule
			
			\cmidrule{1-1} \cmidrule{3-6} \cmidrule{8-11}
			%$64$ & 0.0790 && 0.3670 & 0.7187 & 3.1815 & 1&& -1.0032 & -1.7104 & -21.7969&4\\
			%$128$ & -0.8651&& -9.0714& 297.0923& 46.2143 & 2&& 4.3590& 34.5809& 76.9167&5\\
			%$c$ & 124.27561&& 128.2265& -630.5455& -381.0930& 3&& -121.0518& -137.1210& -220.2500&6\\
			\begin{filecontents*}{energydataOMP.csv}
,ranks,energyKS,energyincreaseKS,workKS,workincreaseKS,energyLF,energyincreaseLF,workLF,workincreaseLF
0,64,1.09$\pm$0.04,1.53x,12405,1.51x,1.11$\pm$0.02,1.57x,12401,1.51x
1,128,1.17$\pm$0.05,1.64x,12536,1.53x,1.19$\pm$0.03,1.68x,12524,1.53x
2,256,1.29$\pm$0.06,1.81x,12809,1.56x,1.34$\pm$0.13,1.89x,12774,1.56x
3,512,1.54$\pm$0.06,2.17x,13374,1.63x,1.48$\pm$0.16,2.08x,13279,1.62x
4,1024,2.21$\pm$0.1,3.12x,14549,1.78x,2.28$\pm$0.15,3.22x,14302,1.75x
\end{filecontents*}
\csvreader[head to column names, late after line=\\,]{energydataOMP.csv}{}% use head of csv as column names
			{\ranks && \workKS & \workincreaseKS & \energyKS & \energyincreaseKS && \workLF & \workincreaseLF & \energyLF & \energyincreaseLF}
			%\bottomrule	
		\end{tabular}
	\end{adjustbox}
	\caption{Work and energy cost in a \emph{full} registration of $\num{4096}$ images on Piz Daint. Results are presented with sample standard deviation and compared against serial execution with $\num{4096} + \num{4095}$ steps and \SI{0.709}{\mega\joule} energy consumption. }
	\label{tab:energy_data}
	
	%\vspace{-10mm}
	%\vspace{-1.5cm}
\end{table}
% PPOPP

% PPOPP - removed 
\subsection{Hierarchical Prefix Scan}
\label{sec:evaluation_hierarchical}

We study the performance effects of the introduction of a hierarchy of parallel workers, replacing the standard set-up of $P$ MPI ranks with $P'$ ranks equipped with $T$ threads each, such that $P' \cdot T = P$ and present in Table~\ref{tab:static_data} a comparison against serial execution and the distributed execution. We observe a significant difference in performance for both the dissemination and \texttt{MPI\_Scan}. On the other hand, we observe performance degradation with the Ladner--Fischer in the sweet spot on 1024 cores. Hierarchical prefix scan leads to performance improvements thanks to a decreased cost of the inter-node synchronization in global phase, except of a single outlier with Ladner--Fischer scan on 1024 cores.
%
%An interesting case is the Ladner-Fischer prefix scan on 1024 cores.
There, the efficiency of a pure MPI solution suddenly increases to the point where it outperforms the dissemination prefix scan by over 60\% and introducing hierarchy leads to performance degradation. We conclude that this specific setup is a sweetspot for the Ladner--Fischer scan where the global scan phase performs very well without significant delays.

%\vspace{-9mm}

\subsection{Work and Energy}
\label{sec:evaluation_energy}

%In Section~\ref{sec:prefix_scan}, we discussed the work-depth relation of prefix scans for which the reduction in depth has to be compensated by an increase in performed work, measured in the number of applications of the operator.
Prefix scan has to compensate reduction in depth by an increase in work. When combined with the load imbalance that further decreases parallel efficiency, the question has to be asked: how much more energy is consumed to reduce the time of processing microscopy images? To find an answer, we measure the energy cost of \textit{full} image registration with preprocessing step on the Piz Daint system with Cray Resource Utilization Reporting (RUR)~\cite{crayrur}. A special consideration has been given for baseline measurements of serial execution because Cray RUR statistics are collected at the node-level only. We allocate the entire node with $12$ MPI ranks computing independently a serial registration and obtain a mean from the total energy cost.
%Since technical limitations prevent us from measuring energy cost of running a job on a single core, we allocate the entire node with $12$ MPI ranks, we let each rank compute independently a serial registration and compute a mean from the total energy cost.}

We estimate work increase on executions with $P$ ranks or with $P'$ ranks and $T$ threads. The work performed in each allocation is computed as a sum of (a) $N$ preprocessing steps, (b) $N \cdot P$ or $N \cdot P' \cdot T$ steps of first phase, (c) global scan and (d) third phase with $N$ steps. In addition, work--stealing jobs include the local scan step over available threads. Results are presented in Table~\ref{tab:energy_data}. Applying our hierarchical and dynamic parallelization decreases the energy consumption up to $2.23x$ and $1.93x$ times when using the dissemination and Ladner--Fischer prefix scan, respectively. This result is higher than the provided speedup since in addition to decreased allocation time, we perform less inter-node communication. The different increases in work and energy cost can be explained by the imbalanced workload causing a drop in parallel efficiency. Such effect is more visible on larger allocations since more time is spent on idle waiting and synchronization.

\begin{figure}[htb]
	\centering
	
	\resizebox*{0.9\width}{0.8\totalheight}{
		\subfloat[The weak scaling of prefix scan image registration.]{
			\includegraphics[width=\dimexpr0.49\textwidth]{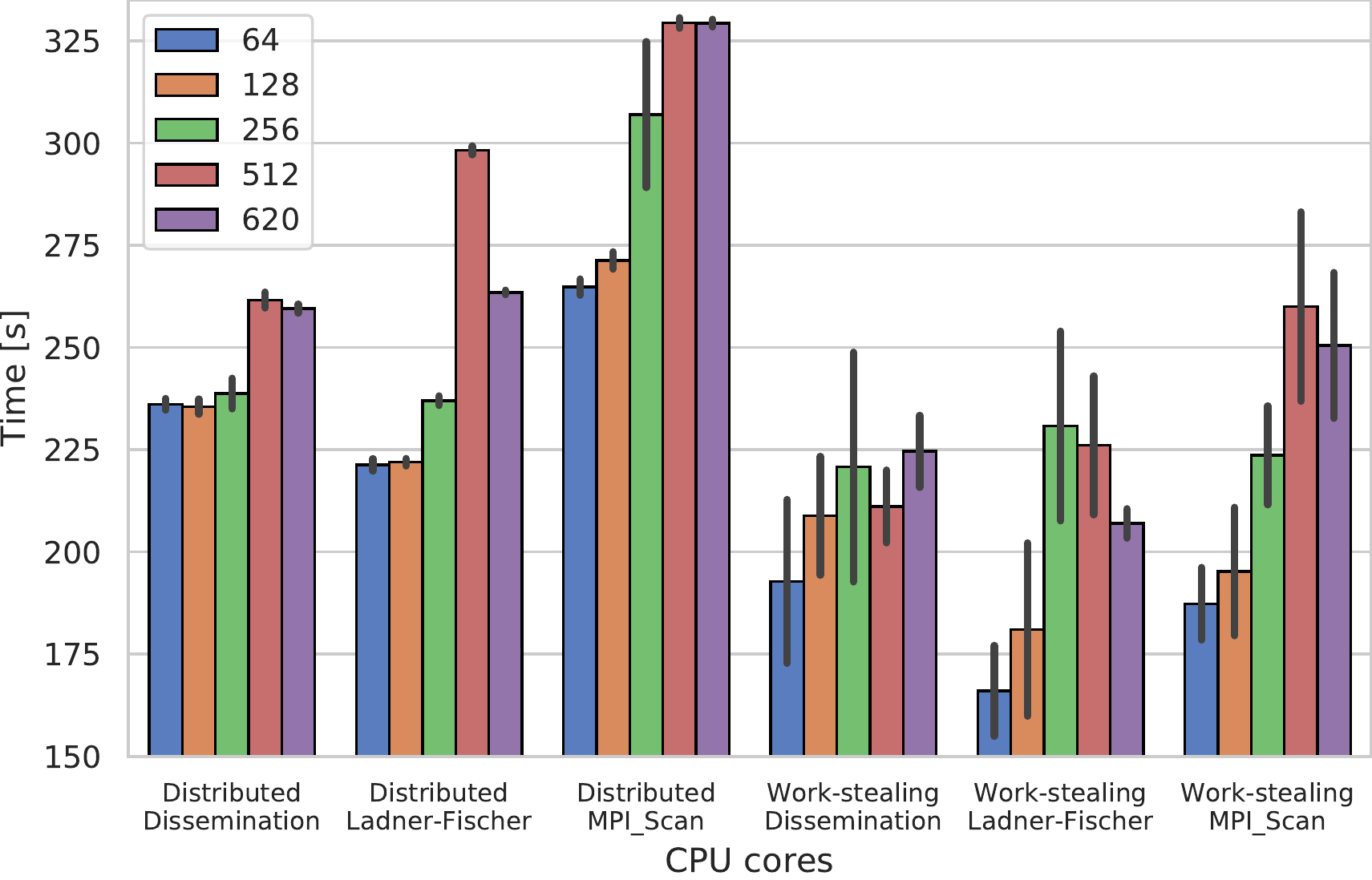} \label{fig:weak_scan}}
	}
	%\hfill
	\newline
	
	\resizebox*{0.9\width}{0.8\totalheight}{
		\subfloat[The weak scaling of \emph{full} image registration with preprocessing.]{
			\includegraphics[width=\dimexpr0.49\textwidth]{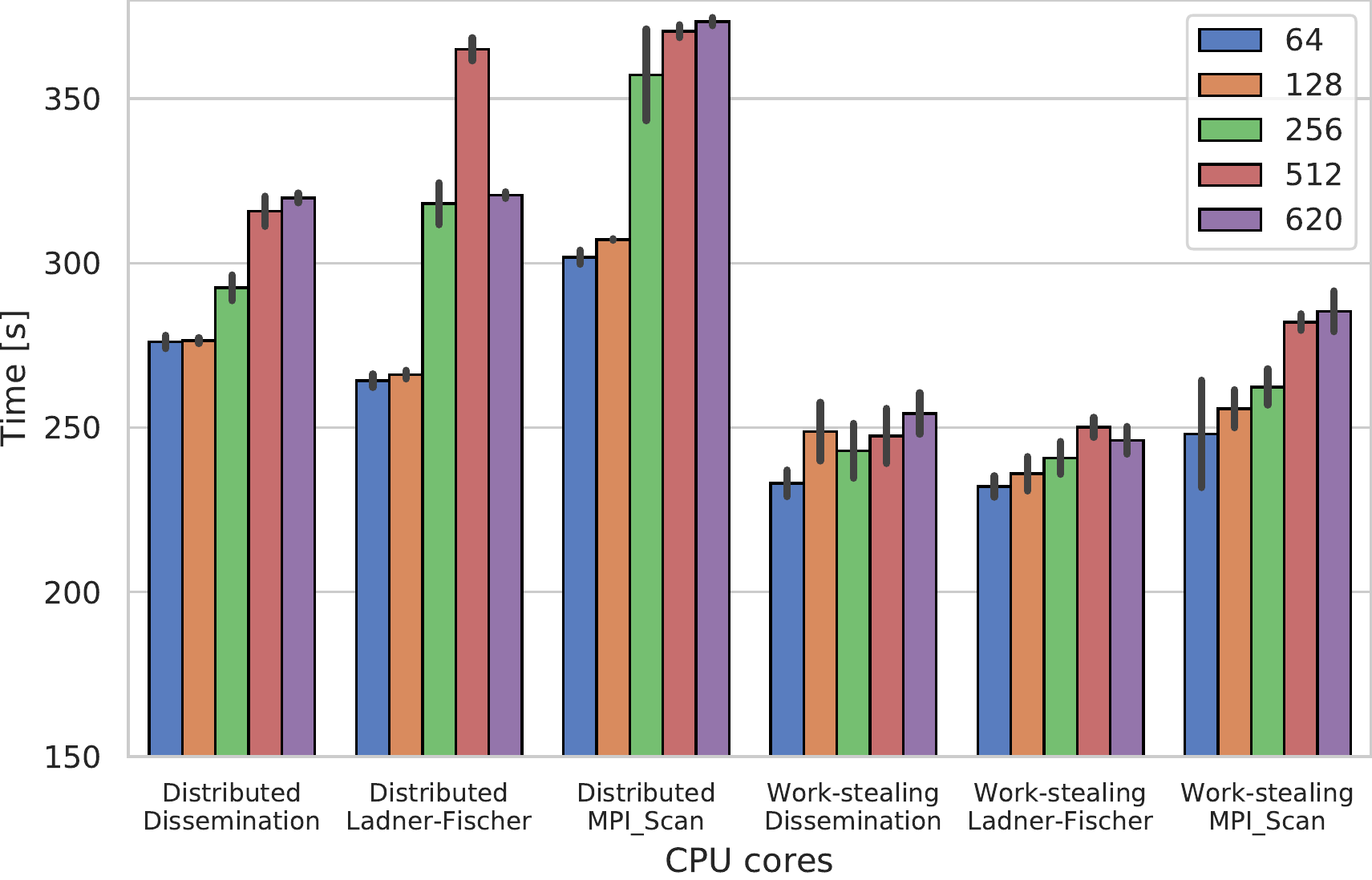} \label{fig:weak_full}}
	}
	%\hfill
	%\vspace{-3mm}
	\caption{The weak scaling of image registration on Ivy Bridge. Figures begin at value 150 for better readability.} 
	%\vspace{-6mm}
\end{figure}

%\vspace{-3mm}
\subsection{Weak Scaling}
\label{sec:evaluation_weak_scaling}

An efficient weak scaling is an important goal since it allows to utilize additional hardware to process longer sequences of microscopy frames while keeping the analysis time practical. When increasing compute resources by the same factor as the problem size, the execution time cannot stay constant due to the logarithmic factor associated with prefix scan: $D_{DS}(k \cdot N, k \cdot P) = D_{DS}(N, P) + C_{1} \log_2{k}$.	
%PPOPP
%According to the depth of parallel prefix scan, when increasing compute resources by the same factor as the problem size, the execution time cannot stay constant due to the logarithmic factor associated with prefix scan: $D_{DS}(k \cdot N, k \cdot P) = D_{DS}(N, P) + C_{1} \log_2{k}$.	

We analyze the weak scaling for the scenario of 8 images per rank and scale it from 64 to 620 ranks on the Ivy Bridge system. Figures~\ref{fig:weak_scan} and~\ref{fig:weak_full} present results for the prefix scan and the full registration procedure, respectively. While we observe an increased execution time for static prefix scan algorithms, the work-stealing procedure helps to mitigate the effects of a logarithmic increase in depth. This benefit is especially visible in the full registration, where for both dissemination and Ladner--Fischer prefix scan the dynamically balanced version scales to a larger number of cores with a minor change in execution time.
%PPOPP
  	%There, the Ladner-Fischer prefix scan tends to deliver slightly better performance than dissemination prefix scan.

\section{Conclusions}
\label{sec:conclusions}
In this paper, we proposed a prefix scan parallelization strategy for the registration of a long
series of electron microscopy images. This work is first to consider a prefix scan problem
where the optimization focus moves from communication latency to a computationally expensive
and highly load-imbalanced operator. To overcome scaling difficulties and slowdowns of prefix scans on imbalanced computations, 
we apply the work--efficient Ladner--Fischer prefix scan
%PPOPP
%to a distributed computation
and provide a novel node-local
work-stealing procedure that can be applied to any prefix scan parallel computation.
We show that work stealing improves the performance of prefix scan on imbalanced workloads up to 2.1x times
while decreasing energy consumption up to 2.23x times.
As a result, an analysis of arbitrarily long microscopy series is now possible thanks 
to a dynamic prefix scan strategy that keeps scaling with increasing hardware resources.

\ifCLASSOPTIONcompsoc
  \section*{Acknowledgments}
\else
  \section*{Acknowledgment}
\fi
The authors would like to thank Professor Sarah Haigh, University of Manchester, for
providing the TEM data.
We would also like to acknowledge the following organizations for providing us
with access to their supercomputers: Swiss National Supercomputing Centre (CSCS),
and the Aachen Institute for Advanced Study in Computational Engineering Science (AICES).

%\ifCLASSOPTIONcaptionsoff
%  \newpage
%\fi
\bibliographystyle{IEEEtran}
\bibliography{paper}

\begin{IEEEbiographynophoto}{Marcin Copik}
is a PhD student at ETH Zurich. His research interests include high-performance computing, serverless computing and performance modeling.
\end{IEEEbiographynophoto}

\begin{IEEEbiographynophoto}{Tobias Grosser}
is an Associate Professor at University of Edinburgh, where he works in the Compiler and Architecture Design Group.
His reasearch interests are in compilation, programming language design, and effective performance programming.
\end{IEEEbiographynophoto}

%\begin{IEEEbiographynophoto}{\scriptsize Torsten Hoefler}\scriptsize
\begin{IEEEbiographynophoto}{Torsten Hoefler} 
  is a Professor at ETH Zurich, where he leads the Scalable Parallel Computing
  Lab. His research aims at understanding performance of parallel computing
  systems ranging from parallel computer architecture through parallel programming
  to parallel algorithms.
\end{IEEEbiographynophoto}

\begin{IEEEbiographynophoto}{Paolo Bientinesi}
is a Professor in High-Performance Computing at Umeå University and the director of High Performance Computing Center North (HPC2N).
His research interests are in numerical linear algebra, tensor operations, performance modelling \& prediction, computer music, and the automatic generation of algorithms and code.
\end{IEEEbiographynophoto}

\begin{IEEEbiographynophoto}{Benjamin Berkels}
is a Juniorprofessor for Mathematical Image and Signal Processing at AICES, RWTH Aachen.
His research interests include variational and joint methods for image registration and segmentation,
and medical image processing.
\end{IEEEbiographynophoto}

\end{document}